# Financial market with no riskless (safe) asset

**Svetlozar (Zari) T. Rachev*** and **Frank J. Fabozzi**
GlimmAnalytics EDHEC Business School

December 5, 2016

**ABSTRACT**

We study markets with no riskless (safe) asset. We derive the corresponding Black-Scholes-Merton option pricing equations for markets where there are only risky assets which have the following price dynamics: (i) continuous diffusions; (ii) jump-diffusions; (iii) diffusions with stochastic volatilities, and; (iv) geometric fractional Brownian and Rosenblatt motions. No arbitrage and market completeness conditions are derived in all four cases.

*Contact information: GlimmAnalytics 121 Thomson Street, Port Jefferson, NY 11777,email:rachev@glimmanalytics.com.

There is an ongoing controversy – the safe asset controversy -- that in real-world financial markets there is a shortage or absence of a riskless asset offering a meaningful positive riskless rate. There are several definitions of what a safe asset is. In his review of the history and economic function of safe assets, Gorton (2016) defines a safe asset as "an asset that is (almost always) valued at face value without expensive and prolonged analysis." Safe assets play several critical roles in a financial market. First, safe assets are used by certain financial entities to satisfy regulatory requirements. Second, they are used as a pricing benchmark. Third, they are used as collateral in financial transactions. Finally, the development of asset pricing theory and derivatives pricing relies on the existence of a safe or riskless asset. However, as Peter Fisher, former Under Secretary of the U.S. Treasury and currently director of BlackRock, Inc., correctly states in an article entitled about the meaning of the riskless rate: "The idea of risk-free sovereign bonds is best thought of as an oxymoron or as an anomaly of recent history. It is not a useful, necessary or an enduring feature of the financial landscape."

There are several studies that have focused on the challenges for monetary policy and global financial stability when there is a shortage of a safe asset (see, for example, Eggertsson and Krugman (2012), Gourinchas and Jeanne (2012), Caballero and Farhi (2013), and Aoki, Nakajima and Nikolov (2014)). Our focus in this paper is on asset pricing when either there is no meaningful riskless asset offering a positive interest rate or if for some reason market participants such as investors and traders elect not to invest in a riskless asset. One of the earlier investigations of the role of a riskless asset in financial theory is by Black (1972) who confirmed that the capital asset pricing model (CAPM) holds even in the absence of a riskless asset (Black's zero-beta CAPM).

Equilibrium models with riskless assets have been extensively studied, see for example Nielsen (1990) Allingham (1991), Konno and Shirakawa (1995), Sun and Yang (2003), and Munk (2006). Our approach to modeling financial markets with no riskless asset is different. We either (i) derive the riskless asset as a perpetual option on the set of risky assets, so that an extended market with a newly created riskless asset provides no arbitrage opportunities and is complete, or (ii) show cases of a complete market where the introduction of the riskless asset leads to arbitrage opportunities, and for those markets the role of a "riskless asset" is played by assets with a stochastic trend.

The paper is organized as follows. A summary of results on option pricing in no-arbitrage complete markets with two risky assets only is provided in the next section. The results are not novel, rather they serve as an illustration of our methodology in the sections that follow. In Section II we derive the multivariate Black-Scholes-Merton (BSM) option pricing formula for perpetual derivatives in markets with multiple risky assets. We do not assume that the riskless asset is initially present in the market. Rather, we construct the price dynamics of the riskless asset from the price dynamics of risky assets. In Section II and the subsequent sections, the riskless asset is derived as a perpetual derivative of all risky assets in the market. Together, the risky assets and the newly introduced riskless asset form a no-arbitrage complete market. In Sections III and IV we extend our results from sections II to the case of markets with prices processes following jump-diffusions and stochastic volatility. In Section V we extend the results from Section II to fractional markets. What is of considerable interest in that section is that while in classical dynamic asset pricing theory the riskless asset always exists, in fractal markets, in contrast, the introduction of a riskless asset leads immediately to pure arbitrage opportunities. This should have been clear, as it is well-known that fractional Brownian Motion (FBM) is not a semi-martingale. Furthermore, when FBM is persistent (i.e., it exhibits long-range dependence), the trader using assets with dynamics following a fractional geometric Brownian motion is, as matter of fact, a clairvoyant. The existence of a riskless asset will allow the trader to apply pure arbitrage strategies. Amazingly, there is a vast literature on markets with price processes following geometric Brownian motion and the riskless asset attempting to fix this obvious fundamental internal defect of the fractional market by introducing a riskless asset. In Section V, we introduce the fractional stochastic safe asset as an alternative to the classical riskless asset. As a result, the fractional market is now arbitrage free and complete. We note that in fractional markets, only perpetual derivatives can be traded. In Section VI we extend the results from Section V to fractional markets with multiple assets. In Section VII we extend the results from Section V to non-Gaussian fractional markets. In the final section, we summarize our findings. All proofs are given in the Appendix.

# I. Market with Two Perfectly Correlated Risky Assets and No Riskless Asset

We start this section with the simplest case of a market with two risky assets (designated by $\mathbb{S}$ and $\mathbb{V}$) with price dynamics following perfectly positively correlated geometric Brownian motions (GBMs). The price dynamics for $\mathbb{S}$ and $\mathbb{V}$ are respectively:

For $\mathbb{S}$: $dS_t = \mu S_t dt + \sigma S_t dB_t, t \geq 0,\ S_0 > 0, \mu > 0, \sigma > 0$ (1)

For $\mathbb{V}$: $dV_t = \mu_V V_t dt + \sigma_V V_t dB_t, t \geq 0,\ S_0 > 0, \mu_V > 0,\ \sigma_V > 0, \sigma_V \neq \sigma.$ (2)

In (1) (resp. (2)) , $\mu$ and $\sigma$ (resp. $\mu_V$ and $\sigma_V$ ) are the instantaneous mean return and the volatility of asset $\mathbb{S}$ (resp. $\mathbb{V}$ ). The Brownian motion, denoted by $B_t$, generates a stochastic basis $(\Omega, \mathcal{F}, \{\mathcal{F}_t\}_{t\geq 0}, \mathbb{P})$ representing the natural world on which the price processes $S_t, t \geq 0$, and $V_t, t \geq 0$ are defined.

## *A*. Deriving the Riskless Rate

To guarantee that the markets (1) and (2) are free of arbitrage and complete, we search for a unique state-price-deflator $\pi_t, t \geq 0$ on $\mathbb{P}$ with dynamics given by an Itô process:

$$d\pi_t = \mu_t^{(\pi)}\pi_t dt + \sigma_t^{(\pi)}\pi_t dB_t.$$

Then the deflated price processes $S_t^{(\pi)} = S_t\pi_t$ and $V_t^{(\pi)} = V_t\pi_t$ should be $\mathbb{P}$-martingales. This leads to

$$\mu_t^{(\pi)} = \frac{\mu\sigma_V - \mu_V\sigma}{\sigma - \sigma_V}, \qquad \sigma_t^{(\pi)} = -\frac{\mu - \mu_V}{\sigma - \sigma_V}.$$

Because the riskless rate is given by $R = -\mu_t^{(\pi)}$, (see Duffie (2001), Section 6D),

$$R = \frac{\mu_V\sigma - \mu\sigma_V}{\sigma - \sigma_V} \tag{3}$$

Indeed, if $\sigma_V = 0$, then $\mu_V$ will be equal to the riskless rate $r$, and $R = r$. Thus, in (3), $R$ represents the riskless rate generated by the market consisting of publicly traded assets $\mathbb{S}$ and $\mathbb{V}$, and it is not defined by a publicly traded riskless bond. We will show later that by having publicly traded assets $\mathbb{S}$ and $\mathbb{V}$, we can construct an asset $\mathbb{B}$ which is a riskless asset having a riskless rate $R$. If this asset $\mathbb{B}$ is introduced as a publicly traded asset, it can play the role of a riskless bond.

## *B*. Pricing a Perpetual European Contingent Claim

Consider a perpetual European contingent claim (ECC) with price process $Y_t = Y(S_t, V_t), t \geq 0$. .We assume that $Y(x, y), x \geq 0, y \geq 0$ is sufficiently smooth function. We also assume that $R \geq 0$[1]. Then the market-price-of-risk $\vartheta$ has the form

[1] If $R < 0$, we must assume that the assets $\mathbb{S}$ and $\mathbb{V}$ are perfectly negative correlated,

$$\vartheta := \frac{\mu - R}{\sigma} = \frac{\mu_V - R}{\sigma_V} = \frac{\mu - \mu_V}{\sigma - \sigma_V}, \tag{4}$$

which again shows that $R$ in (3) can be viewed as the riskless rate. The next proposition represents another proof that $R$ is indeed the riskless rate.

*PROPOSITION 1 (Black-Scholes-Merton[2] equation for market with no riskless bank account): The price process $Y_t = Y(S_t, V_t)$ at $t \in [0, T]$ for the ECC satisfies the following partial differential equation (PDE):*

$$\frac{\partial Y(x,y)}{\partial x} Rx + \frac{\partial Y(x,y)}{\partial y} Ry - RY(x,y) + \frac{1}{2}\frac{\partial^2 Y(x,y)}{\partial x^2}\sigma^2 x^2 + \frac{1}{2}\frac{\partial^2 Y(x,y)}{\partial y^2}\sigma_V^2 y^2 +$$

$$+ \frac{\partial^2 Y(x,y)}{\partial x \partial y}\sigma\sigma_V xy = 0 \tag{5}$$

*Proof of Proposition 1:* See the Appendix.

As an application of Proposition 1 consider a perpetual ECC purchased for $S_0^a V_0^b$ at $t = 0$, and can be sold at any time $t > 0$, for the price of $S_t^a V_t^b$,where $a \in R, b \in R$ are some constants. As per Proposition 1, such a contract can be publicly traded in the no-arbitrage market given by (1) and (2), if and only if the constants $a \in R$ and $b \in R$ satisfy the quadratic equation:

$$aR + bR - R + \frac{1}{2}a(a-1)\sigma^2 + \frac{1}{2}b(b-1)\sigma_V^2 + ab\sigma\sigma_V = 0$$

---

that is, we should replace (2) with

$$dV_t = \mu_V V_t dt - v_V V_t dB_t, t \geq 0,\ S_0 > 0, \mu_V > 0,\ v_V > 0, v_V \neq -\sigma.$$

The analysis of the case $R < 0$ is complete analogous to the case of $R > 0$.

[2] See Black and Scholes (1973) and Merton (1973a).

Notice that (5) is an extension of the classical BSM equation. Indeed, let $\sigma_V \downarrow 0$ in (2), then $\mu_V \downarrow r$, where $r$ is the riskless rate. Thus $V_t$ converges to the riskless asset dynamics $\beta_t = \beta_0 e^{rt}, \beta_0 = V_0$. Then setting $\sigma_V = 0, R = r, C(x,t) = Y(x, \beta_0 e^{rt})$ in the PDE (5) we obtain the BSM-equation:

$$\frac{\partial C(x,t)}{\partial t} + \frac{\partial C(x,t)}{\partial x} rx - rC(x,t) + \frac{1}{2}\frac{\partial^2 C(x,t)}{\partial x^2}\sigma^2 x^2 = 0. \qquad (6)$$

## *C*. The Riskless Asset as a Perpetual Derivative

In the next proposition, we will introduce a new asset (designated as $\mathbb{B}$ ) with price process, $\mathcal{b}_t, t \geq 0$, solely determined by $S_t$ and $V_t$, which can be viewed as a proxy for the riskless asset. Define

$$\mathcal{b}_t = \frac{{V_t}^{\mathcal{v}}}{{S_t}^{\mathcal{v}_V}}, \qquad (7)$$

where

$$\mathcal{v} := \frac{\sigma}{(\sigma - \sigma_V)\left[1 + \frac{1}{2}\sigma\sigma_V\left(\frac{1}{R}\right)\right]} \qquad (8)$$

and

$$\mathcal{v}_V := \frac{\sigma_V}{(\sigma - \sigma_V)\left[1 + \frac{1}{2}\sigma\sigma_V\left(\frac{1}{R}\right)\right]} \qquad (9)$$

*PROPOSITION 2: Suppose that together with assets* $\mathbb{S}$ *and* $\mathbb{V}$, *the asset* $\mathbb{B}$ *with price process* $\mathcal{b}_t, t \geq 0$, *is publicly traded. Then*

$(i)$ $\quad \mathcal{b}_t = \mathcal{b}_0 e^{Rt}, t \geq 0;$ *(10)*

$(ii)$ $\frac{S_t}{\mathscr{b}_t}, \frac{V_t}{\mathscr{b}_t},\ t \geq 0$, *are* $\mathbb{Q}$- *martingales, where* $\mathbb{Q}$ *is the equivalent martingale measure (EMM),* $\mathbb{Q}\sim\mathbb{P}$ *on the stochastic basis* $(\Omega, \mathcal{F}, \{\mathcal{F}_t\}_{t\geq 0}, \mathbb{Q})$ *generated by the Brownian motion* $B_t^{\mathbb{Q}} t \geq 0$, *and on* $\mathbb{P}$,

$$B_t^{\mathbb{Q}} = B_t + \vartheta t, \vartheta = \frac{\mu - R}{\sigma} \qquad (11)$$

$(iii)$ *The market with tradable assets* $\mathbb{S}$, $\mathbb{V}$, *and* $\mathbb{B}$ *is free of arbitrages and complete.*

*Proof of Proposition 2:* See the Appendix.

*COROLLARY 1: For every* $t \in [0, T)$,

$$Y_t = Y(S_t, V_t) = e^{-R(T-t)}\mathbb{E}^{\mathbb{Q}}(\mathcal{G}(S_T, V_T) |\, \mathcal{F}_t). \qquad (12)$$

*Proof of Corollary 1.* Per Proposition 2, the market with assets $\mathbb{S}$, $\mathbb{V}$, and $\mathbb{B}$ is complete, and thus, $\frac{Y_t}{\mathscr{b}_t}, t \geq 0$, is a $\mathbb{Q}$- martingale. This proves (12).

## *D*. The Option Pricing Model

We continue with a simple demonstration of Corollary 1 using Kim-Stoyanov-Rachev-Fabozzi (KSRF) option pricing model, referred to as the multi-purpose binomial option pricing model proposed by Kim at al.(2016). Consider the KSRF stock price model:

$$\left(S_{(k+1)\Delta t}, V_{(k+1)\Delta t}\right)$$

$$= \begin{cases} \left(S^{+}_{(k+1)\Delta t}, V^{+}_{(k+1)\Delta t}\right) = \\ = \left(S_{k\Delta t}\left(1 + \mu\Delta t + \sqrt{\frac{1-p_{\Delta t}}{p_{\Delta t}}}\sigma\sqrt{\Delta t}\right), V_{k\Delta t}\left(1 + \mu_V\Delta t + \sqrt{\frac{1-p_{\Delta t}}{p_{\Delta t}}}\sigma_V\sqrt{\Delta t}\right)\right) \text{with prob. } p_{\Delta t} \\ and \\ \left(S^{-}_{(k+1)\Delta t}, V^{-}_{(k+1)\Delta t}\right) = \\ = \left(S_{k\Delta t}\left(1 + \mu\Delta t - \sqrt{\frac{p_{\Delta t}}{1-p_{\Delta t}}}\sigma\sqrt{\Delta t}\right), V_{k\Delta t}\left(1 + \mu_V\Delta t - \sqrt{\frac{p_{\Delta t}}{1-p_{\Delta t}}}\sigma_V\sqrt{\Delta t}\right)\right) \text{with prob. } 1 - p_{\Delta t} \end{cases}$$

$k = 0,1,\dots,n\Delta t = T$. Here $S_{k\Delta t}$ and $V_{k\Delta t}$ are the prices of the assets $\mathbb{S}$ and $\mathbb{V}$ at $k\Delta t$. Then, *for every fixed* $p_{\Delta t} \in (0,1)$, this binomial tree generates a discrete price process which converges weakly as $n \uparrow \infty$, to the pair of GBMs given by (1) and (2).

Suppose that the option $Y_t = Y(S_t, V_t)$ is risky; that is, it has two possible outcomes (it is not a riskless bond). The trader (designated as ℶ ) takes a short position in the option contract. At $t_k = k\Delta t$ , ℶ forms a portfolio

$$P_{t_k} = Y_{t_k} - \Delta^{(S,k)}S_{t_k} - \Delta^{(V,k)}V_{t_k}.$$

At $t_{k+1} = (k+1)\Delta t$, ℶ$'s$ portfolio has two possible outcomes:

$$P^{(+)}_{t_{k+1}} = Y^{(+)}_{t_{k+1}} - \Delta^{(S,k)}S^{(+)}_{t_{k+1}} - \Delta^{(V,k)}V^{(+)}_{t_{k+1}} \text{ and } P^{(-)}_{t_{k+1}} = Y^{(-)}_{t_{k+1}} - \Delta^{(S,k)}S^{(-)}_{t_{k+1}} - \Delta^{(V,k)}V^{(-)}_{t_{k+1}}.$$

ℶ choses $\Delta^{(S,k)}$ and $\Delta^{(V,k)}$, so that $P^{(+)}_{t_{k+1}} = P^{(-)}_{t_{k+1}} = 0$. That is,

$$\Delta^{(S,k)} = \frac{Y^{(+)}_{t_{k+1}}V^{(-)}_{t_{k+1}} - Y^{(-)}_{t_{k+1}}V^{(+)}_{t_{k+1}}}{S^{(+)}_{t_{k+1}}V^{(-)}_{t_{k+1}} - S^{(-)}_{t_{k+1}}V^{(+)}_{t_{k+1}}} \text{ and } \Delta^{(V,k)} = \frac{Y^{(-)}_{t_{k+1}}S^{(+)}_{t_{k+1}} - Y^{(+)}_{t_{k+1}}S^{(-)}_{t_{k+1}}}{S^{(+)}_{t_{k+1}}V^{(-)}_{t_{k+1}} - S^{(-)}_{t_{k+1}}V^{(+)}_{t_{k+1}}}.$$

Therefore, $P_{t_{k+1}} = 0$, and thus to $P_{t_k} = 0$. As a result, we derive the risk-neutral derivative dynamics given by $Y_{t_k} = \Delta^{(S,k)}S_{t_k} + \Delta^{(V,k)}V_{t_k} = q_{k\Delta t}Y^{+}_{t_{k+1}} + (1 - q_{k\Delta t})Y^{-}_{t_{k+1}}$, where $q_{k\Delta t} =$ $p_{\Delta t} - \vartheta\sqrt{p_{\Delta t}(1-p_{\Delta t})}\sqrt{\Delta t}$ as required for $q_{k\Delta t}$ to be risk-neutral probability, see KSRF (2016), Section 3.2.

## II. Markets with No Riskless Asset: The Case When Asset Prices Follow Correlated GBMs

Consider now a market with $N$ risky assets (designated as $\mathbb{S}^{(j)}, j = 1, \dots, N, N \geq 2$,) with price dynamics following $N$ correlated GBMs:

$$dS_t^{(j)} = \mu^{(j)} S_t^{(j)} dt + S_t^{(j)} \sum_{k=1}^{N-1} \sigma^{(j,k)}\, dB_t^{(k)}, t \geq 0, j = 1, \dots, N, \tag{13}$$

where , $S_0^{(j)} > 0, \mu^{(j)} > 0, \sigma^{(j,k)} > 0$.

### *A*. Deriving the Riskless Rate

The $N$-dimensional price process $\mathcal{S}_t = (S_t^{(1)}, \dots, S_t^{(N)}), t \geq 0,$ is defined on a stochastic basis $(\Omega, \mathcal{F}, \{\mathcal{F}_t\}_{t\geq 0}, \mathbb{P})$[3] , representing the natural world. To guarantee that the market (13) is free of arbitrage opportunities, we search for a unique state-price deflator $\pi_t, t \geq 0$, on $\mathbb{P}$ with dynamics given by the Itô process: $d\pi_t = \mu_t^{(\pi)} \pi_t dt + \sum_{k=1}^{N-1} \sigma^{(\pi,k)} \pi_t\, dB_t^{(k)}$. The existence and uniqueness of $\pi_t$ is equivalent to the requirement that the following linear system

$$\mu_t^{(\pi)} + \mu^{(j)} + \sum_{k=1}^{N-1} \sigma^{(j,k)}\, \sigma^{(\pi,k)} = 0, j = 1, \dots, N$$

has a unique solution. Then we can define riskless rate $R$ to be $R = -\mu_t^{(\pi)}$.

Assume that the price dynamics in (13) is such that the $N \times N$ matrix

[3] $(\Omega, \mathcal{F}, \{\mathcal{F}_t\}_{t\geq 0}, \mathbb{P})$ is generated by the independent Brownian motions $(B_t^{(1)}, \dots, B_t^{(N-1)}), t \geq 0$ .

$$\Phi := \begin{bmatrix} 1 & -\sigma^{(1,1)} & \dots & -\sigma^{(1,N-1)} \\ 1 & -\sigma^{(2,1)} & \dots & -\sigma^{(2,N-1)} \\ \vdots & \vdots & & \vdots \\ 1 & -\sigma^{(N,1)} & \dots & -\sigma^{(N,N-1)} \end{bmatrix}$$

is of full rank and let $det\Phi$ be the determinant of $\Phi$. Then the riskless rate $R$ is determined by

$$R = -\mu_t^{(\pi)} = \frac{det\Phi^{(\mathrm{R})}}{det\Phi} \tag{14}$$

where

$$\Phi^{(\mathrm{R})} = \begin{bmatrix} \mu^{(1)} & -\sigma^{(1,1)} & \dots & -\sigma^{(1,N-1)} \\ \mu^{(2)} & -\sigma^{(2,1)} & \dots & -\sigma^{(2,N-1)} \\ \vdots & \vdots & & \vdots \\ \mu^{(\mathrm{N})} & -\sigma^{(N,1)} & \dots & -\sigma^{(N,N-1)} \end{bmatrix}.$$

An equivalent way of deriving $R$-value can be done by nothing that the vector of market-price-of-risk $\Theta = \left(\theta^{(1)}, \dots, \theta^{(N-1)}\right)^T$, and the riskless rate $R$, should satisfy the system of $N$ equations:

$$\Sigma^{(\mathrm{j})}\Theta = \mathrm{M}^{(j)} - R\mathcal{I}^{(N-1)}, j = 1, \dots, N \tag{15}$$

where

$$\Sigma^{(\mathrm{j})} := \begin{bmatrix} \sigma^{(1,1)} & \sigma^{(1,2)} & & \sigma^{(1,N-1)} & \sigma^{(1,N)} \\ \dots & \dots & & \dots & \dots \\ \sigma^{(j-1,1)} & \sigma^{(j-1,2)} & & \sigma^{(j-1,N-1)} & \sigma^{(j-1,N)} \\ \sigma^{(j+1,1)} & \sigma^{(j+1,2)} & \dots & \sigma^{(j+1,N-1)} & \sigma^{(j+1,N)} \\ \dots & \dots & & \dots & \dots \\ \sigma^{(N,1)} & \sigma^{(N,2)} & & \sigma^{(N,N-1)} & \sigma^{(N,N)} \end{bmatrix},$$

$\mathrm{M}^{(j)} \coloneqq \left(\mu^{(1)}, \dots, \mu^{(\mathrm{j}-1)}, \mu^{(\mathrm{j})}, \dots, \mu^{(\mathrm{N})}\right)^T$, and $\mathcal{I}^{(N-1)} := (1, \dots, 1)^T$ is the unit vector in $R^{N-1}$. Because we have required that the market with securities $\mathbb{S}^{(j)}, j = 1, \dots, N$ , is arbitrage free and complete, then the model coefficients in (13) should be such that $\Theta$ and $R$ are uniquely determined by (15).

The explicit representation for $R$ in (14) is indeed quite complex. For $N = 3$,

$$R = \frac{\left\{\begin{matrix} \mu^{(1)}\left(\sigma^{(1,2)}\sigma^{(2,3)} - \sigma^{(2,2)}\sigma^{(1,3)}\right) - \mu^{(2)}\left(\sigma^{(1,1)}\sigma^{(2,3)} - \sigma^{(2,1)}\sigma^{(1,3)}\right) + \\ +\mu^{(3)}\left(\sigma^{(1,1)}\sigma^{(2,2)} - \sigma^{(2,1)}\sigma^{(1,2)}\right) \end{matrix}\right\}}{\sigma^{(1,1)}\left(\sigma^{(2,2)} - \sigma^{(2,3)}\right) - \sigma^{(1,2)}\left(\sigma^{(2,1)} - \sigma^{(2,3)}\right) + \sigma^{(1,3)}\left(\sigma^{(2,1)} - \sigma^{(2,2)}\right)}.$$

## *B*. Pricing a Perpetual ECC

An alternative proof that $R$ given by (14) is the riskless rate is given by the following proposition. Consider a new asset, a perpetual ECC (designated as $\mathbb{G}$) with price process $\mathrm{G_t} = g(\mathcal{S}_t), t \geq 0$ .We assume that $g(\mathbb{x}), \mathbb{x} = (\mathrm{x}^{(1)}, \dots, \mathrm{x}^{(\mathrm{N})}) \in R_+^N$ ,t is a sufficiently smooth function.

*PROPOSITION 3: (BSM-PDE for markets with no riskless asset): The price process* $\mathrm{G_t} = g(\mathcal{S}_t), t \geq 0$ *for the ECC satisfies the following PDE:*

$$\sum_{j=1}^{N} \frac{\partial g(\mathbb{x})}{\partial x^{(j)}} R\, x^{(j)} - Rg(\mathbb{x}) + \frac{1}{2}\sum_{j=1}^{N} \frac{\partial g^2(\mathbb{x})}{\partial x^{(j)^2}} x^{(j)^2} \sigma^{(j)^2} +$$

$$+ \sum_{i=1}^{N} \sum_{j=i+1}^{\mathrm{N}} \frac{\partial g^2(\mathbb{x})}{\partial x^{(i)} \partial x^{(j)}} \sum_{k=1}^{N-1} \sigma^{(i,k)} \sigma^{(j,k)} = 0.$$

*Proof of Proposition 3:* See the Appendix.

Consider a perpetual ECC which bought at $t = 0$ for $\prod_{j=1}^{N} \left(S_0^{(j)}\right)^{a^{(j)}}$ and can be sold for $\prod_{j=1}^{N} \left(S_t^{(j)}\right)^{a^{(j)}}$ at any time $t > 0$. Such a ECC can be publicly traded in the market represented by (13) if and only if the constants $a^{(j)} \in R, j = 1, \dots, N$ satisfy the equation:

$$R\sum_{j=1}^{N} a^{(j)} - R + \frac{1}{2}\sum_{j=1}^{N}\left(a^{(j)}\left(a^{(j)}-1\right){\sigma^{(j)}}^{2}\right) + \sum_{i=1}^{N}\sum_{j=i+1}^{\mathrm{N}} a^{(i)}a^{(j)}\sum_{k=1}^{N-1}\sigma^{(i,k)}\sigma^{(j,k)} = 0.$$

## $C$. The Riskless Asset as a Perpetual Derivative

How can the market represented by (3) create a riskless asset with riskless rate $R$ which if it is publicly traded can be used as riskless asset? From (13),

$$dlogS_t^{(j)} = \left(\mu^{(j)} - \frac{1}{2}\sum_{k=1}^{N-1}{\sigma^{(j,k)}}^{2}\right)dt + \sum_{k=1}^{N-1}\sigma^{(j,k)}\,dB_t^{(k)}, j = 1,..,N.$$

We search for constants $\chi^{(j)}, j = 1, \dots, N$ such that $\mathscr{b}_t, t \geq 0$ is the riskless asset:

$$dlog\mathscr{b}_t = Rdt = \sum_{j=1}^{N}\chi^{(j)}\,dlogS_t^{(j)} =$$

$$= \sum_{j=1}^{N}\chi^{(j)}\left\{\left(\mu^{(j)} - \frac{1}{2}\sum_{k=1}^{N-1}{\sigma^{(j,k)}}^{2}\right)dt + \sum_{k=1}^{N-1}\sigma^{(j,k)}\,dB_t^{(k)}\right\}.$$

That is,

$$\sum_{j=1}^{N}\left(\mu^{(j)} - \frac{1}{2}\sum_{k=1}^{N-1}{\sigma^{(j,k)}}^{2}\right)\chi^{(j)} = R, \quad \sum_{j=1}^{N}\left(\sum_{k=1}^{N-1}\sigma^{(j,k)}\right)\chi^{(j)} = 0. \qquad (16)$$

The unique solution $\left(\chi^{(1)}, \dots, \chi^{(N)}\right)$ of the linear system (16) provides the form of the riskless price process:

$$\mathscr{b}_t = \prod_{j=1}^{N}\left(S_t^{(j)}\right)^{\chi^{(j)}}, t \geq 0. \qquad (17)$$

If the perpetual asset $\mathbb{B}$ with price process $\mathscr{b}_t, t \geq 0$, given by (17), is publicly traded, then it will serve the role of a riskless bond in the market. The market with assets $\mathbb{S}^{(j)}, j = 1, \dots, N$, and $\mathbb{B}$ is now free of arbitrage opportunities and complete.

## III. Market with No Riskless Assets: The Risky Asset Prices Follow Correlated Jump-Diffusions

Consider now Merton's (1976) (see also Runggaldier (2003)) jump-diffusion model with three assets $\mathbb{S}^{(j)}, j = 1,2,3$ with price processes:

$$\frac{dS_t^{(j)}}{S_{t-}^{(j)}} = \mu_t^{(j)} dt + \sigma_t^{(j)} dB_t + \gamma_t^{(j)} dN_t, t \geq 0, j = 1, 2, 3, \tag{18}$$

where $S_0^{(j)} > 0, \mu^{(j)} \in \mathrm{R}, \sigma^{(j)} > 0, \gamma_t^{(j)} \in R, j = 1,2,3$. The triplet of price processes $\left(S_t^{(1)}, S_t^{(2)}, S_t^{(3)}\right), t \geq 0$, is defined on a stochastic basis $(\Omega, \mathcal{F}, \{\mathcal{F}_t\}_{t\geq 0}, \mathbb{P})$, representing the natural world. The basis $(\Omega, \mathcal{F}, \{\mathcal{F}_t\}_{t\geq 0}, \mathbb{P})$ is generated by the Brownian motion $B_t, t \geq 0$ and a non-homogeneous Poisson process $N_t, t \geq 0$, with intensity $\lambda_t > 0, t \geq 0$. Denote by $M_t = N_t - \int_0^t \lambda_u du, t \geq 0$, the martingale corresponding to $N_t, t \geq 0$.

### *A*. Deriving the Riskless Rate

From (18), it follows that

$$\frac{dS_t^{(i)}}{S_{t-}^{(j)}} = (\mu_t^{(j)} + \gamma_t^{(j)} \lambda_t) dt + \sigma_t^{(j)} dB_t + \gamma_t^{(j)} dM_t, t \geq 0, j = 1,2,3. \tag{19}$$

Under the EMM $\mathbb{Q} \sim \mathbb{P}$,

$$\frac{dS_t^{(i)}}{S_{t-}^{(j)}} = \left(\mu_t^{(j)} + \sigma_t^{(j)} \vartheta_t + \gamma_t^{(j)} \eta_t \lambda_t\right) dt + \sigma_t^{(j)} dB_t^{\mathbb{Q}} + \gamma_t^{(j)} dM_t^{\mathbb{Q}},$$

where $B_t^{\mathbb{Q}}$ is a Brownian motion on $\mathbb{Q}$ and $M_t^{\mathbb{Q}}$ is the Poisson martingale on $\mathbb{Q}$. $\mathbb{P}$, $B_t^{\mathbb{Q}}$ and $M_t^{\mathbb{Q}}$ are determined by the market-price-of-risk densities $\vartheta_s$ and $\lambda_s(1-\eta_s)$:

$$B_t^{\mathbb{Q}} = B_t + \int_0^t \vartheta_s ds\,, M_t^{\mathbb{Q}} = M_t + \int_0^t \lambda_s(1-\eta_s)ds.$$

Then the riskless rate $R_t$ should satisfy the linear system:

$$R_t - \sigma_t^{(j)}\vartheta_t - \lambda_t\gamma_t^{(j)}\eta_t = \mu_t^{(j)}, j = 1,2,3.$$

That is,

$$R_t = \frac{\begin{vmatrix} \mu_t^{(1)} & -\sigma_t^{(1)} & -\lambda_t\gamma_t^{(1)} \\ \mu_t^{(2)} & -\sigma_t^{(2)} & -\lambda_t\gamma_t^{(2)} \\ \mu_t^{(3)} & -\sigma_t^{(3)} & -\lambda_t\gamma_t^{(3)} \end{vmatrix}}{\begin{vmatrix} 1 & -\sigma_t^{(1)} & -\lambda_t\gamma_t^{(1)} \\ 1 & -\sigma_t^{(2)} & -\lambda_t\gamma_t^{(2)} \\ 1 & -\sigma_t^{(3)} & -\lambda_t\gamma_t^{(3)} \end{vmatrix}} \qquad (20)$$

### $B$. The Riskless Asset as a Perpetual Derivative

Let $\mathscr{b}_t = \mathscr{b}_0 e^{\int_0^t R_s ds}, t \geq 0$. Then $\frac{S_t^{(1)}}{\mathscr{b}_t}, \frac{S_t^{(2)}}{\mathscr{b}_t}, \frac{S_t^{(3)}}{\mathscr{b}_t}$ are $\mathbb{Q}$-martingales. If $G_t = g\left(S_t^{(1)}, S_t^{(2)}, S_t^{(3)}, t\right)$ is the price process of a ECC $\mathbb{G}$ with terminal value, $H_T = H\left(S_T^{(1)}, S_T^{(2)}, S_T^{(3)}, T\right)$ then $G_t = e^{\int_t^T R_s ds}\, \mathbb{E}_t^{\mathbb{Q}} H_T$.

Let $\tau_n, n = 1,2, \ldots$ be the jump times of the Poisson process $N_t, t \geq 0$. Then by (18),

$$S_t^{(j)} = S_0^{(j)} e^{\int_0^t \left(\mu_s^{(j)} - \frac{\sigma_s^{(j)^2}}{2}\right) ds + \int_0^t \sigma_s^{(j)} dB_s} \prod_{n=1}^{N_t} \left(1 + \gamma_{\tau_n}^{(j)}\right), j = 1,2,3.$$

It is now clear that in this general setting a construction of the riskless asset (designated as $\mathbb{B}$) with price dynamics $b_t = b_0 e^{\int_0^t R_s ds}, t \geq 0$ in a manner similar to (17) is not possible. However, suppose the jump-diffusion market consists of basic instruments $\mathbb{S}^{(j,B)}, j = 1,2,3$, with price processes

$$\frac{dS_t^{(1,B)}}{S_{t-}^{(1)}} = \mu^{(1)}dt + \sigma^{(1)}dB_t + \gamma dN_t, t \geq 0,$$

$$\frac{dS_t^{(2,B)}}{S_{t-}^{(2,B)}} = \mu^{(2)}dt + \sigma^{(2)}dB_t t \geq 0,.$$

$$\frac{dS_t^{(3,,B)}}{S_{t-}^{(3)}} = \mu^{(3)}dt + \gamma dN_t, t \geq 0. \tag{21}$$

Then the riskless rate is given by

$$R = \frac{\mu^{(2)}\sigma^{(1)} - \sigma^{(2)}\mu^{(1)} + \mu^{(3)}\sigma^{(2)}}{\sigma^{(1)}} \tag{22}$$

and

$$b_t = e^{Rt} = \left(\frac{S_t^{(1)}}{S_0^{(1)}}\right)^{\sigma^{(2)}\Re} \left(\frac{S_t^{(2)}}{S_0^{(2)}}\right)^{-\sigma^{(1)}\Re} \left(\frac{S_t^{(3)}}{S_0^{(3)}}\right)^{-\sigma^{(2)}\Re}, t \geq 0, \tag{23}$$

where $\Re := \frac{R}{\mu^{(1)}\sigma^{(2)} - \mu^{(3)}\sigma^{(2)} - \mu^{(3)}\sigma^{(1)} - \sigma^{(2)}\frac{\sigma^{(1)^2}}{2} + \sigma^{(1)}\frac{\sigma^{(2)^2}}{2}}$.

Assuming that security $\mathbb{B}$ with price process (23) is publicly traded, then $\mathbb{B}$ can serve the role of a riskless asset for the market $\mathbb{S}^{(j,B)}, j = 1,2,3$.

## *C*. Pricing a Perpetual ECC

Consider a perpetual ECC $g\left(\mathcal{S}_t^{(B)}\right), t \geq 0,\ \mathcal{S}_t^{(B)} = \left(S_t^{(1,B)}, S_t^{(2,B)}, S_t^{(3,B)})\right),$ where $g(\mathbb{x}), \mathbb{x} = \left(\mathrm{x}^{(1)}, \mathrm{x}^{(2)}, x^{(3)}\right) \in R_+^3,$ is a sufficiently smooth function.

*PROPOSITION 4: (BSM-PDE for jump-diffusion markets with no riskless asset): The price process $G_t = g(\mathcal{S}), t \geq 0$ for the ECC satisfies the following PDE:*

$$R\frac{\partial g(\mathbb{x})}{\partial x^{(1)}}x^{(1)} + R\frac{\partial g(\mathbb{x})}{\partial x^{(2)}}x^{(2)} + R\frac{\partial g(\mathbb{x})}{\partial x^{(3)}}x^3 - Rg(\mathbb{x}) +$$

$$+\frac{1}{2}\frac{\partial^2 g(\mathbb{x})}{\partial {x^{(1)}}^2}\left(\sigma^{(1)}x^{(1)}\right)^2 + \frac{1}{2}\frac{\partial^2 g(\mathbb{x})}{\partial {x^{(2)}}^2}\left(\sigma^{(2)}x^{(2)}\right)^2 + \frac{\partial g(\mathbb{x})}{\partial x^{(1)}\partial x^{(2)}}\sigma^{(1)}\sigma^{(2)}x^{(1)}x^{(2)} +$$

$$+(R-\mu^{(3)})\gamma\left\{\frac{1}{2}\frac{\partial^2 g(\mathbb{x})}{\partial {x^{(1)}}^2}{x^{(1)}}^2 + \frac{1}{2}\frac{\partial^2 g(\mathbb{x})}{\partial {x^{(3)}}^2}{x^{(2)}}^2 + \frac{\partial g(\mathbb{x})}{\partial x^{(1)}\partial x^{(3)}}x^{(1)}x^{(2)}\right\} = 0$$

*Proof of Proposition 4:* See the Appendix.

Note that if in (21), $\gamma = 0$, then

$$\mu^{(3)} = R = \frac{\mu^{(2)}\sigma^{(1)} - \sigma^{(2)}\mu^{(1)} + R\sigma^{(2)}}{\sigma^{(1)}} = \frac{\mu^{(2)}\sigma^{(1)} - \sigma^{(2)}\mu^{(1)}}{\sigma^{(1)} - \sigma^{(2)}},$$

and, as expected, Proposition 4 becomes Proposition 1.

## IV. Market with no Riskless Assets: Risky Asset Prices Follow Correlated Price Processes with Stochastic Volatility

Consider now a market with price processes with stochastic volatilities. It consists of four traded assets[4]: (i) two stocks $\mathbb{S}^{(j)}, j = 1{,}2$ with price processes

[4] This is a version of the classical stochastic volatility model, see Heston, (1993) and Fouque, Papanicolaou and Sircar (2000)

$$\frac{dS_t^{(j)}}{S_t^{(j)}} = \mu_t^{(j)} dt + g(v_t^{(j)}) dB_t, t \geq 0, j = 1,2, \tag{24}$$

and, (ii) and their volatilities $\mathbb{v}^{(j)}, j = 1,2$ with dynamics given by

$$\frac{dv_t^{(j)}}{v_t^{(j)}} = \alpha^{(j)} dt + \beta^{(j)} dB_t^{(j)}, t \geq 0, \; dB_t dB_t^{(j)} = \rho^{(j)} dt \, , dB_t^{(1)} dB_t^{(2)} = \rho dt, j = 1,2 \, . \tag{25}$$

The function $g(x), x > 0$, is strictly increasing and sufficiently smooth . As examples for $g(x)$, one can consider $g(x) = x^{\nu}, \nu \in (0,1)$, and $g(x) = \log(x)$.

## *A*. Deriving the riskless rate

We require that the stock-market $(\mathbb{S}^{(1)}, \mathbb{S}^{(2)})$ is characterized by no arbitrage opportunities and complete. This model requirement is equivalent to the existence of a unique state-price-deflator

$$\frac{d\pi_t}{\pi_t} = \mu_t^{(\pi)} dt + \sigma_t^{(\pi)} dB_t + \gamma_t^{(1,\pi)} dB_t^{(1)} + \gamma_t^{(2,\pi)} dB_t^{(2)},$$

such that

$$\mu_t^{(j)} + \mu_t^{(\pi)} + \left(\sigma_t^{(\pi)} + \rho^{(1)} \gamma_t^{(1,\pi)} + \rho^{(2)} \gamma_t^{(2,\pi)}\right) g\left(v_t^{(j)}\right) = 0, j = 1,2.$$

Then the security with return $R = -\mu_t^{(\pi)}$ is riskless, and the riskless rate $R$ is given by

$$R = \frac{\mu_t^{(2)} g\left(v_t^{(1)}\right) - \mu_t^{(1)} g\left(v_t^{(2)}\right)}{g\left(v_t^{(1)}\right) - g\left(v_t^{(2)}\right)}.$$

## *B*. Pricing Perpetual ECC

Consider a perpetual derivative $g(\mathcal{S}_t, \boldsymbol{v}_t), t \geq 0, \mathcal{S} = \left(S_t^{(1)}, S_t^{(2)}, v_t^{(1)}, v_t^{(2)})\right)$, where $g(\mathbb{x}, \mathbb{y}), \mathbb{x} =$

$(\mathrm{x}^{(1)}, \mathrm{x}^{(2)}, y^{(1)}, y^{(2)}) \in R_+^4$ is a sufficiently smooth function.

*PROPOSITION 5: (BSM-PDE for stochastic volatility markets with no riskless asset): The price process $G_t = g(\mathcal{S}), t \geq 0$ for the ECC satisfies the following PDE:*

$$R\frac{\partial g(\mathbb{x},\mathbb{y})}{\partial x^{(1)}}x^{(1)} + R\frac{\partial g(\mathbb{x},\mathbb{y})}{\partial x^{(2)}}x^{(2)} + R\frac{\partial g(\mathbb{x},\mathbb{y})}{\partial y^{(1)}}y^{(1)} + R\frac{\partial g(\mathbb{x},\mathbb{y})}{\partial y^{(2)}}y^{(2)} - Rg(\mathbb{x},\mathbb{y}) +$$

$$+\frac{1}{2}\frac{\partial^2 g(\mathbb{x},\mathbb{y})}{\partial {x^{(1)}}^2}g\left(y^{(1)}\right)^2 {x^{(1)}}^2 + \frac{1}{2}\frac{\partial g(\mathcal{S}_t,\boldsymbol{v}_t)}{\partial x^{(2)}}g\left(y^{(2)}\right)^2 {x^{(2)}}^2 + \frac{1}{2}\frac{\partial^2 g(\mathcal{S}_t,\boldsymbol{v}_t)}{\partial {y^{(1)}}^2}\beta^{(1)}{y^{(1)}}^2 +$$

$$+\frac{1}{2}\frac{\partial g(\mathcal{S}_t,\boldsymbol{v}_t)}{\partial y^{(2)}}\beta^2 {y^{(2)}}^2 + \frac{\partial^2 g(\mathcal{S}_t,\boldsymbol{v}_t)}{\partial x^{(1)}\partial x^{(2)}}x^{(1)}x^{(2)}\, g(y^{(1)})g\left(y^{(2)}\right) +$$

$$+\frac{\partial^2 g(\mathcal{S}_t,\boldsymbol{v}_t)}{\partial x^{(1)}\partial y^{(1)}}x^{(1)}y^{(1)}\rho^{(1)}g\left(y^{(1)}\right)\beta^{(1)} + \frac{\partial^2 g(\mathcal{S}_t,\boldsymbol{v}_t)}{\partial x^{(1)}\partial y^{(2)}}x^{(1)}y^{(2)}\rho^{(2)}g\left(y^{(1)}\right)\beta^{(2)} +$$

$$+\frac{\partial^2 g(\mathcal{S}_t,\boldsymbol{v}_t)}{\partial x^{(2)}\partial y^{(1)}}x^{(2)}y^{(1)}\rho^{(1)}g\left(y^{(2)}\right)\beta^{(1)} + \frac{\partial^2 g(\mathcal{S}_t,\boldsymbol{v}_t)}{\partial x^{(2)}\partial y^{(2)}}x^{(2)}y^{(2)}\rho^{(2)}g\left(y^{(2)}\right)\beta^{(2)}$$

$$+\frac{\partial^2 g(\mathcal{S}_t,\boldsymbol{v}_t)}{\partial y^{(1)}\partial y^{(2)}}v_t^{(1)}v_t^{(2)}\rho\beta^{(1)}\beta^{(2)}.$$

*Proof of Proposition 5:* The proof is similar to the one of Proposition 4 and omitted.

## V. Fractional Market with No Riskless Assets when Risky Asset Prices Follow Perfectly Correlated GFBMs

Fractional markets are extremely popular in academic research and at the same time very controversial.[5] Fractional markets with a riskless asset are not arbitrage free. The asset prices in fractional markets are persistent, that is, the traders in those markets are informed about the upcoming future price movements. We show that such fractional markets are equally informed; that is, they all share the same information in a publicly traded fractional market, there is no arbitrage opportunity and even more the market is complete.

[5] Peters (1994), Mandelbrot (2008) , Rostek (2009), Panas and Ninni (2010, Anderson and Noss (2013), and Dar, Bhanja and Tiwari (2015).

### *A*. Fractional Markets with a Riskless Asset Admit Arbitrage Opportunities

In this section we provide an illustration of fractional markets with no riskless asset. What is obvious is that the presence of a riskless asset in persistent fractional markets will allow the trader in fractional markets to borrow funds at the riskless rate and invest them in arbitrage strategies freely available in the fractional market. Thus, for a fractional market to exist, there should not be a riskless asset.

The centerpiece of the fractional markets is the fractional Brownian motion (FBM[6]) $B_t^{(H)}, t \geq 0$ as a model for market uncertainty.[7] Formally, FBM $B_t^{(H)}, t \geq 0$ with Hurst index $H \in (0,1)$ is defined as a stochastic integral with respect to the BM $B_t, t \geq 0$,

$$B_t^{(H)} = c_H \int_{-\infty}^{\infty} \left( [max(0, t-s)]^{H-\frac{1}{2}} - [\max(0, -s)]^{H-\frac{1}{2}} \right) dB_s \qquad (26)$$

where $c_H = \sqrt{\frac{2H\Gamma(\frac{3}{2}-H)}{\Gamma(\frac{1}{2}+H)\Gamma(2-2H)}}$. FBM is a Gaussian process with

$$\mathbb{E}B_t^{(H)} = 0, \mathbb{E}\left(B_t^{(H)} B_s^{(H)}\right) = \frac{1}{2}\{|t|^{2H} + |s|^{2H} - |t-s|^{2H}\} \text{ for all } t \geq 0, s \geq 0.$$

---

[6] FRM was first introduced by Kolmogorov (1940).
[7] Mandelbrot and van Ness (1968) and Øksendal, (2006).

FBM has stationary increments and is self-similar with index $H$; that is for all $a > 0$,

$$B_{at}^{(H)} \triangleq a^H B_t^{(H)}, t \geq 0^8. \quad (27)$$

For $H = \frac{1}{2}, B_t^{(H)}$ becomes a standard Brownian motion. The case $H \in \left(\frac{1}{2}, 1\right)$ is of special interest in the finance literature because it leads to markets with stock prices exhibiting long- range dependence (LRD)[9]. LRD is a main characteristic of markets with high-frequency trading[10].

From now on we shall assume that $H \in \left(\frac{1}{2}, 1\right)$. In this case we deal with (i) a persistent FBM (i.e., a FBM with positively correlated increments) and (ii) a FBM with LRD. For FBM, the LRD property follows from the following observation. Consider the sequence of increments $X_k^{(H)} = B_k^{(H)} - B_{k-1}^{(H)}, k = 1,2 \ldots$ and let $\rho_n^{(H)} = cov\left(X_k^{(H)}, X_{k+n}^{(H)}\right)$ be the autocovariance function. A stationary sequence $X_k^{(H)}$ is said to exhibit LRD if $\lim_{n\uparrow\infty} C\rho_n^{(H)} n^{\alpha} = 1$ for some constants $C > 0$ and $\alpha \in (0,1)$. For $H \in \left(\frac{1}{2}, 1\right)$, the LRD condition is satisfied with $C = \frac{1}{H(2H-1)}$ and $\alpha = 2(H-1)$.

The trajectories of $B_t^{(H)}, t \geq 0$, are almost nowhere differentiable. However, we can choose a version of $B_t^{(H)}, t \geq 0$, such that all sample paths are continuous and even more: trajectories are continuous and Lipschitz-smooth of order $\alpha$, (shortly, $Lip(\alpha)$- trajectories), for all $\alpha \in (\frac{1}{2}, H)$. That is, for all $t \geq 0$ and $s \geq 0$,

$$\left|B_t^{(H)} - B_s^{(H)}\right| \leq C^{(\alpha)}|t-s|^{\alpha},$$

[8] $\triangleq$ stands for equal of all finite distributions.
[9] For a comprehensive study of LRD for stochastic processes, see Samorodnitsky (2016).
[10] See Willinger, Taqqu, andTeverovsky (1999).

for some positive constant $C^{(\alpha)}$. The smoothness of the trajectories, the $B_t^{(H)}, t \geq 0$ is guaranteed by the finiteness of the $p$- variation ( $p = 1/H$) of $B_t^{(H)}$:

$$lim_{n\uparrow\infty} \sum_{i=1,\ldots,n,t_i^{(n)}=i\frac{t}{n}} \left| B_{t_{i+1}^{(n)}}^{(H)} - B_{t_i^{(n)}}^{(H)} \right|^p < \infty.$$

$B_t^{(H)}, t \geq 0$ is neither a semimartingale nor a Markov process, and requires different stochastic calculus. Stochastic calculus with FBM is based on fractional Stratonovich integral (FSI): for a continuous function $f: [0, T] \to R$, the FSI is denoted by $\int_0^T f(s) \circledast dB_s^{(H)}$ and is defined a limit of the Riemann sums[11]

$$\int_0^T f(s) \circledast dB_s^{(H)} = lim_{n\uparrow\infty} \sum_{k=0,,t^{(k)}=\frac{k}{n}T,k=0,\ldots,n}^{n-1} f\left(t^{(k)}\right) \left(B_{t^{(k+1)}}^{(H)} - B_{t^{(k)}}^{(H)}\right) \qquad (27)$$

This implies the following chain of rule of integration: given a sufficiently smooth function $G(x, t), x \in R, t \geq 0$,

$$G\left( B_{t+s}^{(H)}, t + s \right) = G\left( B_t^{(H)}, t \right) + \int_t^{t+s} \frac{\partial G(B_t^{(H)}, u)}{\partial x} \circledast dB_u^{(H)} + \int_t^{t+s} \frac{\partial G(B_t^{(H)}, u)}{\partial u} du,$$

or in differential terms[12]:

---

[11] Because $H \in \left(\frac{1}{2}, 1\right)$, the integral

$$\int_0^T f(s) \circledast dB_s^{(H)} =$$

$$= lim_{n\uparrow\infty} \sum_{k=0,,t^{(k)}=\frac{k}{n}T,k=0,\ldots,n}^{n-1} f\left((1-\delta)t^{(k)} + \delta t^{(k+1)}\right) \left(B_{t^{(k+1)}}^{(H)} - B_{t^{(k)}}^{(H)}\right)$$

has the same value for all $\delta \in [0,1]$.

[12] In the literature two other notations are used instead of $\ldots \circledast dB_u^{(H)}$:

(i) $\ldots . \delta dB_u^{(H)}$;

$(ii)$ $(S)$ $\ldots . dB_u^{(H)}$

$$d\left( B_t^{(H)}, t\right) = \frac{\partial G(B_t^{(H)},t)}{\partial x} \circledast dB_u^{(H)} + \frac{\partial G\left(B_t^{(H)},t\right)}{\partial t} dt. \quad (28)$$

In this section, we study a fractional market with two risky assets (designated by $\mathbb{S}^{(H)}$ and $\mathbb{V}^{(H)}$) with price dynamics following perfectly positively correlated GFBMs). The price dynamics for $\mathbb{S}^{(H)}$ and $\mathbb{V}^{(H)}$ are given by

$$\text{For } \mathbb{S}^{(H)}: \quad S_t = S_0 \, e^{\mu t + \sigma dB_t^{(H)}}, t \geq 0,\ S_0 > 0, \mu > 0, \sigma > 0 \quad (29)$$

$$\text{For } \mathbb{V}^{(H)}: \quad V_t = V_t \, e^{\mu_V t + \sigma_V B_t^{(H)}}, t \geq 0,\ S_0 > 0, \mu_V > 0,\ \sigma_V > 0, \sigma_V \neq \sigma. \quad (30)$$

Applying pathwise integration (28) [13] we obtain the representation of the stocks' dynamics in differential terms:

$$dS_t = \mu S_t dt + \sigma S_t \circledast dB_t^{(H)}, t \geq 0,\ S_0 > 0, \mu > 0, \sigma > 0, \quad (31)$$

$$dV_t = \mu_V V_t dt + \sigma_V V_t \circledast dB_t^{(H)}, t \geq 0,\ S_0 > 0, \mu_V > 0,\ \sigma_V > 0. \quad (32)$$

In (31) (resp. (32)) , $\mu$ and $\sigma$ (resp. $\mu_V$ and $\sigma_V$) are the instantaneous mean return and the volatility of asset $\mathbb{S}^{(H)}$ (resp. $\mathbb{V}^{(H)}$ ). The FBM, $B_t^{(H)}$, generates a stochastic basis $(\Omega, \mathcal{F}, \{\mathcal{F}_t\}_{t\geq 0}, \mathbb{P})$ representing the natural world on which the price processes $S_t, t \geq 0$, and $V_t, t \geq 0$ are defined.

Let us derive the first condition for the fractal market model (26) and (27) to be free of arbitrage opportunities. Consider a self-financing portfolio $P_t = \sigma_V S_t - \sigma V_t$. Then $dP_t = \sigma_V dS_t - \sigma dV_t = R dt$, where $R = \mu\sigma_V - \mu_V \sigma$. If $R \geq 0$, the market $\left(\mathbb{S}^{(H)}, \mathbb{V}^{(H)}\right)$ admits arbitrage opportunity. To see this, let $\mathbb{B}$ be the riskless bond with price process $dP_t = RP_t dt, t \geq 0$, $P_0 > 0$. Assume that $R \geq 0$[14]. Define a self-financing portfolio[15] with $\mathbb{B}$ and $\mathbb{S}^{(H)}$

$$X_t = b_t P_t + c_t S_t, t \geq 0 \tag{33}$$

where $b_t = 1 - \exp\left\{2B_t^{(H)}\right\}$ and $c_t = -2 + 2\exp\left\{2B_t^{(H)}\right\}$. Then $dX_t = b_t dP_t + c_t dS_t$, for all $t \geq 0$. Furthermore, $X_0 = 0$, but for all $t > 0$,

$$X_t = e^{Rt}\left(-1 + \exp\left\{B_t^{(H)}\right\}\right)^2 > 0 \tag{34}$$

with $\mathbb{P}$-probability 1.

[13] Bender, Sottinen, and Valkeila (2008) and Rostek (2009).

*Therefore, the first condition for fractal markets to exist is that no self-financing portfolio with riskless returns should be publicly traded.* This is discouraging as it will mean that in (28) and (29) we should have $\mu = 0$ and $\mu_V = 0$. Thus the model (29) and (30) has a very limited scope (it is a model with zero drift) and we abandon it in the rest of our study.

### *B*. The Introduction of a Fractional Riskless Asset

As seen from (34), stochastic drifts can be (and should be) used to replace the deterministic drifts in (31) and (32) when dealing with fractal markets.

In view of this observation we revise (29) and (30 ) as follows:

$$For\ \ \mathbb{S}^{(H)}:\ S_t = S_0 \exp\left\{\left(B_t^{(H)}\right)^2 \mu + \sigma B_t^{(H)}\right\}, t \geq 0,\ S_0 > 0, \mu > 0, \sigma > 0 \tag{35}$$

---

[14] The case $R < 0$, is considered in the same manner.
[15] See, for example, Rostek (2009, pp. 59-60).

$$For \quad \mathbb{V}^{(H)}: \; V_t = V_0 \exp\left\{\left(B_t^{(H)}\right)^2 \mu_V + \sigma_V B_t^{(H)}\right\}, t \geq 0, \; S_0 > 0, \mu_V > 0, \; \sigma_V > 0. \tag{36}$$

Per (28), we can re-write (35) and (36) in differential terms:

$$\frac{dS_t}{S_t} = \left(2B_t^{(H)}\mu + \sigma\right) * dB_t^{(H)}, \; \frac{dV_t}{V_t} = \left(2B_t^{(H)}\mu_V + \sigma_V\right) * dB_t^{(H)} \tag{37}$$

Now, we must define what will be a fractional riskless asset (or a fractional bond). We assume that in (35) and (36), $\mu\sigma_V \neq \mu_V\,\sigma$. Then we obtain the dynamics of an asset with pure fractional stochastic drift:

$$\mathscr{b}_t^{(H,R)} := \left(\frac{S_t}{S_0}\right)^{\frac{\sigma_V R}{\mu\sigma_V - \mu_V\,\sigma}} \left(\frac{V_t}{V_0}\right)^{-\frac{\sigma R}{\mu\sigma_V - \mu_V\,\sigma}} = e^{\,\mathrm{R}\left(B_t^{(H)}\right)^2}, t \geq 0 \tag{38}$$

The process $\mathscr{b}_t^{(H,R)}, t \geq 0, \xi_0 > 0$, can be viewed as the price dynamics of a fractional riskless asset (designated as $\mathbb{B}^{(H,R)}$) with fractional rate $R$. Note that $\mathbb{E}\left(log\mathscr{b}_t^{(H)}\right) = \mathbb{E}\left(B_t^{(H)}\right)^2 = t^{2H}, H \in \left(\frac{1}{2}, 1\right)$. That is, only in the case of the BM, $B_t = B_t^{(1/2)}, t \geq 0$, we have $\mathbb{E}\left(log\mathscr{b}_t^{(1/2)}\right) = t$. That is, when the Hurst index increases (i.e., when the fractional bond exhibits higher level of LRD), the faster (on average) the stochastic drift evolves. That effect shows again that using a riskless asset in markets with persistent fractional price processes is meaningless, as it will lead to arbitrage opportunities.

## *C*. Pricing Perpetual ECC

For which $R$, the fractional asset (or fractional bond) $\mathbb{B}^{(H,R)}$ together with assets $\mathbb{S}^{(H)}$ and $\mathbb{V}^{(H)}$ form a complete market? As we will show in the next, Proposition 6, we must have

$$R = \frac{\mu_V\sigma - \mu\sigma_V}{\sigma - \sigma_V}, \tag{39}$$

which is the same as in (3) in the case of a BSM-market with no riskless asset.

To prove (39), consider a perpetual ECC, $G_t = g(S_t, V_t)$. Note that in fractional markets, one cannot value ECC with future terminal value. The reason for that is the fact that the ECC-seller (designated by ב) can use an arbitrage strategy. Because the FBM exhibits LRD. ב can forecast the future prices, as ב$'s$ information set includes future price values. Thus, ב can employ arbitrage strategies while replicating the ECC-value until maturity. Meanwhile, the ECC buyer (designated as ד) has already paid for the long position in the ECC contract and can do nothing. ד could only wait for the ECC payoff to be realized at the terminal time (and hope for the best). Meanwhile ב will achieve arbitrage profits, while ב will have none. Therefore, only perpetual ECC contacts can be valued within fractional markets. This will be illustrated in the following proposition.

Consider a perpetual ECC with price process $\mathrm{G_t} = g(S_t, V_t), t \geq 0$. We assume that $g(\mathrm{x}, \mathrm{y}), x \geq 0, y \geq 0$ is sufficiently smooth function.

*PROPOSITION 6: (PDE for fractional markets with no riskless asset): Assume that in (29) and (30),* $\sigma\mu_V \neq \mu\sigma_V$. *Then the price process* $G_t = g(S_t, V_t), t \geq 0$ *for the ECC satisfies the following PDE:*

$$\frac{\partial g(x,y)}{\partial x} x + \frac{\partial g(x,y)}{\partial y} y - g(x,y) = 0$$

*Proof of Proposition 6*: See the Appendix.

As an application of Proposition 6, consider a perpetual ECC which bought at time $t = 0$, for $S_0^a\, V_0^b$, where $a \in R$ and $b \in R$ are some constants. The ECC can be sold for $S_t^a\, V_t^b$, at any time $t > 0$. This ECC can be publicly traded in the fractional market defined by (35) and (36) if and only if $a + b = 1$. This proves (39).

Next, consider the dynamics of the stock price $S_t$, discounted by the stochastic fractional bank:

$$d\frac{S_t}{\mathscr{b}_t^{(H,R)}} = \frac{1}{\mathscr{b}_t^{(H,R)}} S_t \left(\left(2B_t^{(H)}(\mu - \mathrm{R}) + \sigma\right) \circledast dB_t^{(H)}\right).$$

Let $B_t^{(H,\mathbb{Q})} := B_t^{(H)} + \theta {B_t^{(H)}}^2$.We can view $B_t^{(H,\mathbb{Q})}$ as FBM with fractional stochastic drift $\theta {B_t^{(H)}}^2$. Then choosing $\theta = \frac{\mu - \mathrm{R}}{\theta}$ leads to the following analogue of asset valuation in the risk-neutral world:

$$d\frac{S_t}{\mathscr{b}_t^{(H,R)}} = \frac{1}{\mathscr{b}_t^{(H,R)}} S_t \left(\sigma \circledast dB_t^{(H,\mathbb{Q})}\right).$$

## $\boldsymbol{C}$. Pricing Perpetual ECC on Stock-Paying Stochastic Dividend

Suppose the stock $\mathbb{S}^{(H)}$ pays "fractal-dividends" at constant rate $D_y$, that is, its price dynamics is given by

$$\frac{dS_t}{S_t} = \left(2B_t^{(H)}(\mu - D_y) + \sigma\right) \circledast dB_t^{(H)}$$

Consider an ECC $C_t = C\left(S_t, \mathscr{b}_t^{(H,R)}\right)$, where $C(\mathrm{x},\mathrm{y}), x \geq 0, y \geq 0$ is sufficiently smooth function. Using the same arguments as in the proof of Proposition 6, we show that $C(\mathrm{x},\mathrm{y})$ satisfies the PDE:

$$\left(R - D_y\right)\frac{\partial C(x,y)}{\partial x} x + R\frac{\partial C(x,y)}{\partial y} y - RC(x,y) = 0 \qquad (41)$$

Now let us consider the option pricing having as underlying assets the stock $\mathbb{S}^{(H)}$ paying a fractal dividend rate $D_y$, and the riskless asset $\mathbb{B}^{(H,R)}$. Let $\mathbb{C}$ be a perpetual ECC with price process $C_t = C\left( S_t, \mathscr{b}_t^{(H,R)}\right), t \geq 0$, where $C(x, y), x > 0, y > 0$ is a sufficiently smooth function. Suppose the ECC-buyer (designated as ד ) would like to have ECC following the dynamics

$$C_t = C_0 \exp\left\{\left(B_t^{(H)}\right)^2 \mu^{(C)} + \sigma^{(C)} B_t^{(H)}\right\}$$

for a pre-specified fixed *fractal information ratio*:

$$\mathbb{J} = \frac{\mu^{(C)}}{\sigma^{(C)}}.$$

The PDE (41) implies that, for every pair of constants $\alpha$ and $\beta$ such that

$$\left(1 - D_y^{(H)}\right)\alpha + \beta - 1 = 0,\ D_y^{(H)} := \frac{D_y}{R}$$

the ECC-value $C_t$ can be replicated by the price process $S_t^{\alpha} \mathscr{b}_t^{(H,R)^{\beta}}$. As a matter of fact,

$$C_t = C\left( S_t, \mathscr{b}_t^{(H,R)}\right) = S_t^{\alpha} \mathscr{b}_t^{(H,R)^{\beta}} = C_0 \exp\left\{\left(B_t^{(H)}\right)^2 \mu^{(C)} + \sigma^{(C)} B_t^{(H)}\right\}.$$

Thus, the required value for $\alpha$ and $\beta$ are $\alpha = \frac{R}{\mathbb{J}\sigma + \left(1 - D_y^{(H)}\right)R - \mu}$ and $\beta = \frac{\mathbb{J}\sigma - \mu}{\mathbb{J}\sigma + \left(1 - D_y^{(H)}\right)R - \mu}$.The initial value of the ECC-contract is $C_0 = S_0^{\alpha}$. The seller of $\mathbb{C}$ will hedge perfectly the liability (namely, $C_t$ ) at time $t \geq 0$ by keeping the publicly traded security with price $S_t^{\alpha} \mathscr{b}_t^{(H,R)^{\beta}}, t \geq 0$, until the ECC is exercised, and thus the short position in the contract is closed.

## VI. Fractional Market: The Multivariate Case

As became clear from Section V, for fractional markets to exist, there should *not* be a riskless asset. Instead in factional markets one can and should introduce a riskless asset with a fractional stochastic rate. In this section, we will extend the results from the previous section to the multivariate case.

Consider now a market with $N$ risky assets (designated as $\mathbb{S}^{(H,j)}, j = 1, \ldots, N, N \geq 2,$) with price dynamics following $N$ fractional price processes $\mathcal{S}_t = \left(S_t^{(1)}, \ldots, S_t^{(N)}\right)$:

$$S_t^{(j)} = \quad S_0^{(j)} \exp\left\{\left(\sum_{m=1}^{N-1} B_t^{(H,m)}\right)^2 \mu^{(j)} + \sum_{k=1}^{N-1} \sigma^{(j,k)} B_t^{(H,k)}\right\}, t \geq 0, j = 1, \ldots, N, \qquad (41)$$

where $S_0^{(j)} > 0, \mu^{(j)} > 0, \sigma^{(j,k)} > 0$. $\mathcal{S}_t, t \geq 0$, defined on a stochastic basis $(\Omega, \mathcal{F}, \{\mathcal{F}_t\}_{t\geq 0}, \mathbb{P})$[16], representing the natural world. In pathwise differential terms, the price dynamics of $\mathbb{S}^{(H,j)}, j = 1, \ldots, N$ is given by:

$$\frac{dS_t^{(j)}}{S_t^{(j)}} = \sum_{k=1}^{N-1}\left(2\mu^{(j)} \sum_{m=1}^{N-1} B_t^{(H,m)} + \sigma^{(j,k)}\right) \circledast dB_t^{(H,k)}, t \geq 0, j = 1, \ldots, N \qquad (42)$$

## *A*. Pricing Perpetual ECC

Consider a new asset, a perpetual ECC (designated as $\mathbb{G}$) with price process $\mathrm{G}_t = g(\mathcal{S}_t), t \geq 0$ .We assume that $g(\mathbb{x}), \mathbb{x} = (\mathrm{x}^{(1)}, \ldots, \mathrm{x}^{(N)}) \in R_+^N$, is a sufficiently smooth function.

*PROPOSITION 7: (PDE for fractional markets with no riskless asset ): Assume that*

$det\Xi \neq 0$, *where*

[16] $(\Omega, \mathcal{F}, \{\mathcal{F}_t\}_{t\geq 0}, \mathbb{P})$ is generated by the independent fractional Brownian motions $(B_t^{(H,1)}, \ldots, B_t^{(H,N-1)}), t \geq 0$ .

$$\Xi = \begin{bmatrix} \mu^{(1)} & \mu^{(2)} & \dots & \mu^{(N-1)} & \mu^{(N)} \\ \sigma^{(1,1)} & \sigma^{(2,1)} & \dots & \sigma^{(N-1,1)} & \sigma^{(N,1)} \\ \sigma^{(1,2)} & \sigma^{(2,2)} & \dots & \sigma^{(,N-1,2)} & \sigma^{(N,2)} \\ \vdots & \vdots & \ddots & \vdots & \vdots \\ \sigma^{(1,N-1)} & \sigma^{(2,,N-1)} & \dots & \sigma^{(N-1,N-1)} & \sigma^{(N,N-1)} \end{bmatrix} \tag{43}$$

*Then the price process* $G_t = g(\mathcal{S}_t), t \geq 0$ *for the ECC satisfies the following PDE:*

$$\sum_{j=1}^{N} \frac{\partial g(\mathbb{x})}{\partial x^{(j)}} x^{(j)} - g(x, y) = 0 \tag{44}$$

*Proof of Proposition 7:* See the Appendix.

## *B*. The Introduction of Fractional Riskless Asset Rate

Consider a perpetual derivative, $\mathbb{G}^{(a)}, a = (a^{(1)}, \dots, a^{(N)}) \in R^N$ with price process

$$g^{(a)}(\mathcal{S}_t) = \prod_{j=1}^{N} \left(S_t^{(j)}\right)^{a^{(j)}}, a = (a^{(1)}, \dots, a^{(N)}) \tag{45}$$

Then, per Proposition 7, $\mathbb{G}^{(a)}$, can be publicly traded in the market defined by securities as $\mathbb{S}^{(H,j)}, j = 1, \dots, N$ if an only if

$$\sum_{j=1}^{N} a^{(j)} = 1. \tag{46}$$

Next, let us determine the fractional riskless asset (or, the fractional bond for the market with securities $\mathbb{S}^{(H,j)}, j = 1, \dots, N$ ) $\mathbb{B}^{(H)}$, with price dynamics

$$\mathscr{b}_t^{(H)} = \exp\left\{\mathrm{R}^{(\mathrm{H})}\left(\sum_{m=1}^{N-1} B_t^{(H,m)}\right)^2\right\}, t \geq 0, \tag{47}$$

where $\mathrm{R}^{(\mathrm{H})}$ is the instantaneous fractional riskless rate.

From definition (41) for the stock price fractal dynamics, it follows that for every $b^{(j)} \in R, j = 1,..,N,$

$$\prod_{j=1}^{N}\left(\frac{S_t^{(j)}}{S_0^{(j)}}\exp\left\{-\mu^{(j)}\left(\sum_{m=1}^{N-1}B_t^{(H,m)}\right)^2\right\}\right)^{b^{(j)}} = \exp\sum_{k=1}^{N-1}\left(\sum_{j=1}^{N}\sigma^{(j,k)}b^{(j)}\right)B_t^{(H,k)} \ , t \geq 0.$$

Per (46) and (47) we want to choose $b^{(j)}, j = 1,..,N$ so that $\sum_{j=1}^{N} b^{(j)} = 1$ and $\sum_{j=1}^{N}\sigma^{(j,k)}b^{(j)} = 0\,, k = 1,\dots,N-1.$ Assume that $det\Psi \neq 0$, where

$$\Psi = \begin{bmatrix} 1 & & 1 & & 1 \\ \sigma^{(1,1)} & \cdots & \sigma^{(j,1)} & \cdots & \sigma^{(N,1)} \\ \vdots & \ddots & \vdots & \ddots & \vdots \\ \sigma^{(1,N-1)} & & \sigma^{(j,N-1)} & & \sigma^{(N,N-1)} \end{bmatrix}.$$

Then solving the linear system for $b^{(j)}, j = 1,..,N$ leads to $b^{(j)} = \frac{det\Psi^{(j)}}{det\Psi}$, where

$$\Psi^{(j)} = \begin{bmatrix} 1 & & 1 & 1 & 1 & & 1 \\ \sigma^{(1,1)} & \cdots & \sigma^{(j-1,1)} & 0 & \sigma^{(j+1,1)} & \cdots & \sigma^{(N,1)} \\ \vdots & \ddots & & \vdots & & \ddots & \vdots \\ \sigma^{(1,N-1)} & & \sigma^{(j-1,N-1)} & 0 & \sigma^{(j+1,N-1)} & & \sigma^{(N,N-1)} \end{bmatrix}.$$

Thus, we can introduce, the fractional riskless asset $\mathbb{B}^{(H)}$ with price dynamics

$$\mathscr{b}_t^{(H)} := \prod_{j=1}^{N}\left(\frac{S_t^{(j)}}{S_0^{(j)}}\right)^{b^{(j)}} = \exp\left\{R^{(H)}\left(\sum_{m=1}^{N-1}B_t^{(H,m)}\right)^2\right\}, \mathrm{R}^{(\mathrm{H})} = \sum_{j=1}^{N}\mu^{(j)}b^{(j)},$$

and fractional riskless asset rate $\mathrm{R}^{(\mathrm{H})}$. By Proposition 7, the asset $\mathbb{B}^{(H)}$ can be publicly traded within the fractal market $\mathbb{S}^{(H,j)}, j = 1,\dots,N.$ In other words, the extended market $\left(\mathbb{B}^{(H)}, \mathbb{S}^{(H,1)},\dots,\mathbb{S}^{(H,N)}\right)$ is a complete market.

### *C*. Beta Model for Fractional Markets

Denote $\mathbb{X}_t = \left(X_t^{(0)}, X_t^{(1)}, \dots, X_t^{(N)}\right) = \left( \mathscr{b}_t^{(H)}, \mathcal{S}_t\right)$, where $\mathcal{S}_t = \left(S_t^{(1)}, \dots, S_t^{(N)}\right)$, $t \in [0,T]$[17]. Then, on $(\Omega, \mathcal{F}, \{\mathcal{F}_t\}_{t\geq 0}, \mathbb{P})$, the $N+1$- GFBM $\mathbb{X}_t$ has the following dynamics

$$\frac{dX_t^{(j)}}{X_t^{(j)}} = \sum_{k=1}^{N-1}\left(2\mu^{(j)} \sum_{m=1}^{N-1} B_t^{(H,m)} + \sigma^{(j,k)}\right) \circledast dB_t^{(H,k)}, t \geq 0, j = 0, \dots, N \tag{49}$$

where $\mu^{(0)} = R^{(H)}$ is given by (48), and $\sigma^{(0,k)} = 0, k = 1, \dots, N$. A trading strategy[18] is a process $\Theta_t = (\theta_t^{(0)}, \dots, \theta_t^{(N)})$ with continuous trajectories, such that the Stratonovich integral

$$\int_0^t \Theta_t \circledast d\mathbb{X}_t \coloneqq \sum_{j=0}^{N} \theta_t^{(j)} \circledast dX_t^{(j)} =$$

$$= \sum_{k=1}^{N-1} \left\{\left[\sum_{j=0}^{N} 2\mu^{(j)}\theta_t^{(j)} X_t^{(j)}\right] \sum_{m=1}^{N-1} B_t^{(H,m)} + \left[\sum_{j=0}^{N} \sigma^{(j,k)}\, \theta_t^{(j)} X_t^{(j)}\right]\right\} \circledast dB_t^{(H,k)} \tag{50}$$

exist.

The value of the investor's portfolio at time $t \in [0,T]$ (or his/her wealth $W_t$) is given by

$$W_t = \Theta_t \mathbb{X}_t \coloneqq \sum_{j=0}^{N} \theta_t^{(j)} X_t^{(j)}$$

---

[17] $\mathbb{X}_t, t \geq 0$ is defined on the stochastic basis $(\Omega, \mathcal{F}, \{\mathcal{F}_t\}_{t\geq 0}, \mathbb{P})$ generated by the independent fractional Brownian motions $(B_t^{(H,1)}, \dots, B_t^{(H,N-1)}), t \geq 0$ .

[18] We follow the general outline of the proof of Merton's ICAPM provided in Chapter 9, Duffie (2001) in the classical case of Brownian motion, $H = 1/2$. Indeed, because we are dealing with a persistent FBM ($H > \frac{1}{2}$), our result will be strikingly different. No continuity of the classical ICAPM ($H = \frac{1}{2}$) and our ICAPM results as $H \downarrow \frac{1}{2}$ will be observed. This is partly due to the fact that as soon as $H > \frac{1}{2}$, the increments of $B_t^{(H,k)}$ are positively correlated while in the case of BM ($H = \frac{1}{2}$), they are independent.

Then

$$\frac{dW_t}{W_t} = \frac{1}{W_t} \sum_{j=0}^{N} \theta_t^{(j)} dX_t^{(j)} = \frac{1}{W_t} \sum_{j=0}^{N} \theta_t^{(j)} X_t^{(j)} \frac{dX_t^{(j)}}{X_t^{(j)}} = \sum_{j=0}^{N} \varphi_t^{(j)} \frac{dX_t^{(j)}}{X_t^{(j)}}.$$

Hence,

$$\frac{dW_t}{W_t} = \sum_{j=0}^{N} \varphi_t^{(j)} \frac{dX_t^{(j)}}{X_t^{(j)}} = \sum_{j=0}^{N} \varphi_t^{(j)} \sum_{k=1}^{N-1} \left( 2\mu^{(j)} \sum_{m=1}^{N-1} B_t^{(H,m)} + \sigma^{(j,k)} \right) \divideontimes dB_t^{(H,k)} =$$

$$= \sum_{k=1}^{N-1} \left( 2 \left( \sum_{j=0}^{N} \varphi_t^{(j)} \mu^{(j)} \right) \sum_{m=1}^{N-1} B_t^{(H,m)} + \left( \sum_{j=0}^{N} \varphi_t^{(j)} \sigma^{(j,k)} \right) \right) \divideontimes dB_t^{(H,k)}$$

Consider an investor whose portfolio, with price process $P_t$ consisting of only risky assets $\mathbb{S}^{(H,j)}, j = 1, \dots, N, N \geq 2$,

$$\frac{dP_t}{P_t} = \sum_{j=1}^{N} \varphi_t^{(j,P)} \frac{dX_t^{(j)}}{X_t^{(j)}} =$$

$$= \sum_{k=1}^{N-1} \left( 2 \left( \sum_{j=0}^{N} \varphi_t^{(j,P)} \mu^{(j)} \right) \sum_{m=1}^{N-1} B_t^{(H,m)} + \left( \sum_{j=0}^{N} \varphi_t^{(j,P)} \sigma^{(j,k)} \right) \right) \divideontimes dB_t^{(H,k)}$$

with $\sum_{j=1}^{N} \varphi_t^{(j,P)} = 1$. The investor minimizes

$$\sum_{k=1}^{N-1}\left(\sum_{j=1}^{N}\varphi_t^{(j,P)}\,\sigma^{(j,k)}\right)^2$$

subject to

$$\sum_{j=0}^{N}\varphi_t^{(j,P)}\left(\mu^{(j)}-\mu^{(W)}\right)=0,\sum_{j=0}^{N}\varphi_t^{(j,P)}=1,$$

where $\mu^{(W)}$ is a benchmark value. Applying the same equilibrium arguments as in Section 20.4, Varian (1992), and Section 6D, Duffie 2001, leads to the following beta model for fractional markets:

$$\sum_{i=1}^{N}\varphi_t^{(j,P)}\mu^{(i)}-R^{(H)}=\frac{\sum_{i=1}^{N}\sum_{j=1}^{N}\varphi_t^{(j,P)}\varphi_t^{(j,M)}\left(\sum_{k=1}^{N-1}\sigma^{(j,k)}\sigma^{(i,k)}\right)}{\sum_{i=1}^{N}{\varphi_t^{(j,M)}}^2\left(\sum_{k=1}^{N-1}{\sigma^{(j,k)}}^2\right)}\left(\sum_{i=1}^{N}\varphi_t^{(j,M)}\mu^{(i,M)}-R^{(H)}\right)\tag{51}$$

where $M_t$ is the market-value-process with price dynamics given by

$$\frac{dM_t}{M_t}=\sum_{j=1}^{N}\varphi_t^{(j,M)}\frac{dX_t^{(j)}}{X_t^{(j)}}.$$

## VII. Markets with Fractional Rosenblatt Motion

We now extend the results from Section VI to non-Gaussian fractional markets.

### *A*. Definition and Properties of Fractional Rosenblatt Process

Fractional Rosenblatt motion (FRM) is more flexible (but more complex) than the FBM model for describing market uncertainty in fractional markets. FRM is a non-Gaussian process, allowing for a better fit of asset returns to real data. It is defined[19] as weak limit of the sequence of processes[20]

$$Z_t^{(n)} = \frac{\sigma}{n^H} \sum_{j=1}^{\lfloor nt \rfloor} \xi^{(j)}, t \in [0,T], \qquad (52)$$

where $\lfloor a \rfloor$ denotes the integer part of $a$ and

$(R1)$ $\sigma > 0$ is a scale parameter, while $H \in \left(\frac{1}{2}, 1\right)$ is the self-similarity (Hurst) index;

$(R2)$ $\xi^{(j)} = {\eta^{(j)}}^2 - 1$, where $\{\eta^{(j)}, j = 0,1,\dots\}$ is a Gaussian sequence with $\mathbb{E}\eta^{(j)} = 0$, $\mathbb{E}\left({\eta^{(j)}}^2\right) = 1$ and autocovariances $\rho^{(n)} = \mathbb{E}\left(\eta^{(0)}\eta^{n}\right) = (1+n^2)^{-\frac{H-1}{2}}$.

The limiting process is the FRM $\mathcal{R}_t^{(H)}, t \geq 0$. $\mathcal{R}_t^{(H)}$ is a Wiener-Itô stochastic integral with respect to the BM $B_t, t \in [0,T]$:

$$\mathcal{R}_t^{(H)} = C^{(\mathcal{R},H)} \iint_{\{(u,v)\in R^2, u\neq v\}} \left[\int_0^t (s-u)_+^{\frac{H}{2}-1} (s-v)_+^{\frac{H}{2}-1}\right] dB_u dB_v \qquad (53)$$

where $C^{(\mathcal{R},H)} = \frac{\sqrt{\frac{H}{2}(2H-1)}}{B\left(\frac{H}{2}, 1-H\right)}$[21]. The choice of the normalizing constant $C^{(\mathcal{R},H)}$ guarantees that $\mathbb{E}\left(\mathcal{R}_1^{(H)}\right)^2 = 1$.

Some properties of the FRM:

---

[19] See Taqqu (2011) and Torres and Tudor (2009)

[20] The convergence of the sequence of $\{Z_t^{(n)}, n \geq 0\}$ is understood as convergence of all finite distributions of $Z_t^{(n)}$, as $n \uparrow \infty$.

[21] $B(a,b) = \frac{\Gamma(a+b)}{\Gamma(a)\Gamma(b)}$ is the beta function

$\left(\mathcal{R}^{(\mathrm{H})}\mathrm{I}\right)$ $\mathcal{R}_t^{(H)}, t \in [0, T]$, is a stationary non-Gaussian process with mean zero;

$\left(\mathcal{R}^{(\mathrm{H})}\mathrm{II}\right)$ The trajectories of $\mathcal{R}_t^{(H)}, t \in [0, T]$ have continuous and Lipschitz-smooth of order $\alpha < H$;

$\left(\mathcal{R}^{(\mathrm{H})}\mathrm{III}\right)$ The covariance function of $\mathcal{R}_t^{(H)}$ is given by $\mathbb{E}\left(\mathcal{R}_t^{(H)}\mathcal{R}_s^{(H)}\right) = \frac{1}{2}(t^{2H} + s^{2H} - |t - s|^{2H}), s, t \in [0, T]$;

$\left(\mathcal{R}^{(\mathrm{H})}\mathrm{IV}\right)$ $\left(Joseph\ effect^{22} - strong\ positive\ LRD\right)$ The increments of $\mathcal{R}_t^{(H)}$

$$\Delta\mathcal{R}_n^{(H)} = \mathcal{R}_{n+1}^{(H)} - \mathcal{R}_n^{(H)}, n = 1,2, \ldots$$

exhibit LRD, as the autocorrelation of the increments decays very slowly:

$$\lim_{n\uparrow\infty} \mathbb{E}\left[\Delta\mathcal{R}_0^{(H)}\Delta\mathcal{R}_n^{(H)}\right] n^{2-2H} = \sigma^2 H(2H - 1).$$

As a consequence,

$$\sum_{n=0}^{\infty} \mathbb{E}\left[\Delta\mathcal{R}_0^{(H)}\Delta\mathcal{R}_n^{(H)}\right] = \infty$$

[22] Benoit Mandelbrot coined the term referring to Genesis 41,29-30: Seven years of great abundance are coming throughout the land of Egypt, but seven years of famine will follow them. Then all the abundance in Egypt will be forgotten, and the famine will ravage the land.”

$\left(\mathcal{R}^{(\mathrm{H})}\mathrm{V}\right)$ $\left(Noah\ effect^{23} - exceptionally\ large\ values\right)$ $\mathcal{R}_t^{(H)}$ is self-similar of order $H$, that is, $\mathcal{R}_{ct}^{(H)} \triangleq c^H \mathcal{R}_t^{(H)}$, for all $c > 0$, and all $ct, t \in [0, T]$;

($\mathcal{R}^{(\mathrm{H})}$VI) $\mathcal{R}_t^{(H)}$ is nowhere differentiable in mean square sense as :

$$lim_{h \downarrow 0} \mathbb{E}\left[\left(\frac{\mathcal{R}_{t+h}^{(H)} - \mathcal{R}_t^{(H)}}{h}\right)^2\right] = lim_{h \downarrow 0} \mathbb{E}\left(\frac{\mathcal{R}_h^{(H)}}{h}\right)^2 = lim_{h \downarrow 0} t^{2H-2} = \infty;$$

Still one can interpret $\mathcal{R}_t^{(H)}$ as

$$\mathcal{R}_t^{(H)} = C^{(\mathcal{R},H)} \int_0^t \left(\frac{\partial}{\partial s} B_s^{\left(H^{(\mathcal{R})}\right)}\right)^2 ds,$$

where $\frac{\partial}{\partial t} B_t^{\left(H^{(\mathcal{R})}\right)}$ is the "generalized derivative"[24] of FBM $B_t^{\left(H^{(\mathcal{R})}\right)}$ with Hurst index $H^{(\mathcal{R})} = \frac{H+1}{2}$;

($\mathcal{R}^{(\mathrm{H})}$VII) (*Heavy-distributional tails*[25] *and extreme values* [26]). For some constant $\mathbb{C} > 0$,

$$\mathbb{P}\left(\mathcal{R}_1^{(H)} > u\right) \leq \mathbb{C} e^{-\frac{u}{2}}, u \geq 0,$$

and

---

[23] Benoit Mandelbrot coined the term referring to Genesis 7:11;8:1-2, "And the rain was upon the earth forty days and forty nights", describes the nature of the flood waters as a cosmic cataclysm.

[24] This representation can be made precise using generalized functions, see Taqqu (2011).

[25] See Major (2005).

[26] See Albin (1998)/

$$limsup_{u\uparrow\infty}\frac{\mathbb{P}\left(sup_{t\in[0,1]}\mathcal{R}_t^{(H)}>u\right)}{u^{\frac{1}{H}-1}\mathbb{P}\left(\mathcal{R}_1^{(H)}>u\right)}<\infty$$

## *B*. The Fractional Rosenblatt Market

We now introduce a Rosenblatt fractional market with two risky assets (designated by $\mathbb{S}^{(\mathcal{R})}$ and $\mathbb{V}^{(\mathcal{R})}$) with price dynamics following perfectly positively correlated geometric fractional Rosenblatt motions (GFRMs). The price dynamics for $\mathbb{S}^{(\mathcal{R})}$ and $\mathbb{V}^{(\mathcal{R})}$ are given by

For $\mathbb{S}^{(\mathcal{R})}$: $S_t = S_0\, \mathrm{e}^{\mu t+\sigma d\mathcal{R}_t^{(H)}}, t\geq 0,\ S_0>0, \mu>0, \sigma>0$ (54)

For $\mathbb{V}^{(\mathcal{R})}$: $V_t = V_t\, e^{\mu_V t+\sigma_V\mathcal{R}_t^{(H)}}, t\geq 0,\ S_0>0, \mu_V>0,\ \sigma_V>0, \sigma_V\neq\sigma.$ (55)

Notice that a market with two assets: the risky asset $\mathbb{S}^{(\mathcal{R})}$, and (ii) a riskless asset $\mathbb{B}$ with price dynamics $\beta_t=\beta_0\exp\left\{\int_0^t r_s ds\right\}, t\geq 0,$ permits pure arbitrage strategies[27], as it was the case with GFBM. That is why we consider the market with two risky assets $\mathbb{S}^{(\mathcal{R})}$ and $\mathbb{V}^{(\mathcal{R})}$.

Applying pathwise integration (28)[28] we obtain the representation of the stocks' dynamics in differential terms:

$$dS_t = \mu S_t dt + \sigma S_t \circledast d\mathcal{R}_t^{(H)}, t\geq 0,\ S_0>0, \mu>0, \sigma>0 \quad (56)$$

$$dV_t = \mu_V V_t dt + \sigma_V V_t \circledast d\mathcal{R}_t^{(H)}, t\geq 0,\ S_0>0, \mu_V>0,\ \sigma_V>0. \quad (57)$$

In (56) (resp. (57)) , $\mu$ and $\sigma$ (resp. $\mu_V$ and $\sigma_V$ ) are the instantaneous mean return and the volatility of asset $\mathbb{S}^{(H)}$ (resp. $\mathbb{V}^{(H)}$ ). The FRM, $\mathcal{R}_t^{(H)}$ generates a stochastic basis

[27] Examples for such arbitrages are provided in Torres and Tudor (2009).
[28] See Zähle (1998).

$(\Omega, \mathcal{F}, \{\mathcal{F}_t\}_{t\geq 0}, \mathbb{P})$ representing the natural world on which the price processes $S_t, t \geq 0,$ and $V_t, t \geq 0$ are defined.

Then as in (38) we can introduce the set with Rosenblatt fractional rate $\mathrm{R}^{(\mathcal{R})}$ and price dynamics

$$\mathscr{b}_t^{(H,R)} := \left(\frac{S_t}{S_0}\right)^{\frac{\sigma_V R}{\mu\sigma_V - \mu_V \sigma}} \left(\frac{V_t}{V_0}\right)^{-\frac{\sigma R}{\mu\sigma_V - \mu_V \sigma}} = e^{\mathrm{R}^{(\mathcal{R})}\left(\mathcal{R}_t^{(H)}\right)^2}$$

where the stochastic drift $\left(\mathcal{R}_t^{(H)}\right)^2$, has the same behavior as $\left(B_t^{(H)}\right)^2$, as $\mathbb{E}\left(\mathcal{R}_t^{(H)}\right)^2 = \mathbb{E}\left(B_t^{(H)}\right)^2 = t^{2H}, H \in \left(\frac{1}{2}, 1\right)$. Now all results we obtained for FBM can be readily proved for FRM.

As a final remark for this section the result for FRM can be directly extended to even more flexible *fractional Hermite motion*[29] , $\mathcal{H}_t^{(H,k)}, t \geq 0, s$, of order $k = 1,2, \ldots$. For $k = 1$, $\mathcal{H}_t^{(H,k)}$ is a FBM, and for $k = 2$, $\mathcal{H}_t^{(H,2)}$ is FRM. The additional parameter $k$ allows for general slow decay of the autocorrelation of the stock return process viewed as geometric FHM. Fitting fractional Hermite motion to real asset return data is the subject of a different paper.

## VIII. Conclusion

Often the riskless asset is not available for trade, does not exist, is in scarce supply or is not desirable to be traded. In this paper, we study markets with no riskless asset initially introduced in the market. The riskless asset is derived as a perpetual derivative of the risky assets determining the market. We developed this approach in three markets, each with no riskless asset: (i) markets with multiple risky assets having continuous diffusion price dynamics; (ii) markets with prices processes following jump diffusions, and; (iii) markets with prices following diffusions with stochastic volatilities. Introducing a completely novel approach we study fractional markets, where the existence of a riskless asset leads to arbitrage opportunities. We derive a fractional riskless asset (fractional bond) as a perpetual derivative of the assets in the fractional market. Our fractional market models are free of arbitrage opportunities and complete. In all considered cases, we derive BSM-type equations for perpetual derivatives.

[29] See Fauth and Tudor (2016).

## APPENDIX

### *Proof of Proposition 1*

Consider a European contingent claim (ECC) with price process $\mathrm{Y_t} = Y(S_t, V_t)$ at $t \in [0, T]$, maturity $T$, terminal value $\mathrm{Y_T} = \mathcal{G}(S_T, V_T)$, and price dynamics given by the Itô process:

$$d\mathrm{Y_t} = dY(S_t, V_t) = \begin{pmatrix} \frac{\partial Y(S_t,V_t)}{\partial x} \mu S_t + \frac{\partial Y(S_t,V_t)}{\partial y} \mu_V V_t + \\ + \frac{1}{2}\frac{\partial^2 Y(S_t,V_t)}{\partial x^2} \sigma^2 S_t^2 + \frac{1}{2}\frac{\partial^2 Y(S_t,V_t)}{\partial y^2} \sigma_V^2 V_t^2 + \\ + \frac{\partial^2 Y(S_t,V_t)}{\partial x \partial y} \sigma S_t \sigma_V V_t \end{pmatrix} dt +$$

$$+ \left[\frac{\partial Y(S_t,V_t)}{\partial x} \sigma S_t + \frac{\partial Y(S_t,V_t)}{\partial y} \sigma_V V_t\right] dB_t \,.$$

Consider the self-financing portfolio:

$$Y(S_t, V_t) = a_t S_t + b_t V_t = (a_t \mu S_t + b_t \mu_V V_t)dt + (a_t \sigma S_t + b_t \sigma_V V_t) dB_t.$$

Comparing the terms with $dY(S_t, V_t)$ leads to:

$$a_t S_t = \frac{1}{\sigma - \sigma_V}\left(\frac{\partial Y(S_t, V_t)}{\partial x} \sigma S_t + \frac{\partial Y(S_t, V_t)}{\partial y} \sigma_V V_t - \sigma_V Y(S_t, V_t)\right)$$

and $b_t V_t = \frac{1}{\sigma - \sigma_V}\left(Y(S_t, V_t)\sigma - \frac{\partial Y(S_t,V_t)}{\partial x} \sigma S_t - \frac{\partial Y(S_t,V_t)}{\partial y} \sigma_V V_t\right)$. Next, applying

$$\frac{\partial Y(S_t, V_t)}{\partial x}\mu S_t + \frac{\partial Y(S_t, V_t)}{\partial y}\mu_V V_t + \frac{1}{2}\frac{\partial^2 Y(S_t, V_t)}{\partial x^2}\sigma^2 S_t^2 + \frac{1}{2}\frac{\partial^2 Y(S_t, V_t)}{\partial y^2}\sigma_V^2 V_t^2$$

$$+ \frac{\partial^2 Y(S_t, V_t)}{\partial x \partial y}\sigma S_t \sigma_V V_t = a_t \mu S_t + b_t \mu_V V_t$$

and setting $S_t = x, V_t = y$, leads to

$$\frac{\partial Y(x, y)}{\partial x}Rx + \frac{\partial Y(x, y)}{\partial y}Ry - RY(x, y) + \frac{1}{2}\frac{\partial^2 Y(x, y)}{\partial x^2}\sigma^2 x^2 + \frac{1}{2}\frac{\partial^2 Y(x, y)}{\partial y^2}\sigma_V^2 y^2$$

$$+ \frac{\partial^2 Y(x, y)}{\partial x \partial y}\sigma \sigma_V xy = 0.$$

Suppose $\sigma_V \downarrow 0$ in $dV_t = \mu_V V_t dt + \sigma_V V_t dB_t, t \geq 0,\ S_0 > 0, \mu_V > 0, \sigma_V > 0$. Then, $\mu_V \downarrow r$, where $r$ is the riskless rate and thus $V_t$ converges to the riskless bond dynamics $\beta_t = \beta_0 e^{rt}, \beta_0 = V_0$. Then setting $\sigma_V = 0, R = r,\ C(x, t) = Y(x, \beta_0 e^{rt})$ leads to $\frac{\partial C(x,t)}{\partial t} = \frac{\partial Y(x,y)}{\partial y}\frac{\partial \beta_0 e^{rt}}{\partial t} = \frac{\partial Y(x,y)}{\partial y}\beta_0 e^{rt} r = R\frac{\partial Y(x,)}{\partial y}y$, and thus we obtain the BSM-equation:

$$\frac{\partial C(x, t)}{\partial t} + \frac{\partial C(x, t)}{\partial x}rx - rC(x, t) + \frac{1}{2}\frac{\partial^2 C(x, t)}{\partial x^2}\sigma^2 x^2 = 0.$$

This completes the proof of Proposition 1.

***Proof of Proposition 2***

From (1) and (2) it follows that

$$S_t = S_0 e^{\left(\mu - \frac{1}{2}\sigma^2\right)t + \sigma B_t}, V_t = V_0 e^{\left(\mu_V - \frac{1}{2}\sigma_V^2\right)t + \sigma_V B_t},$$

Thus,

$$\frac{1}{\sigma}\log\left(\frac{S_t}{S_0}\right)-\frac{1}{\sigma_V}\log\left(\frac{V_t}{V_0}\right)=\left\{\frac{\mu-\frac{1}{2}\sigma^2}{\sigma}-\frac{\mu_V-\frac{1}{2}\sigma_V^2}{\sigma_V}\right\}t$$

and $\frac{\frac{1}{\sigma}\log\left(\frac{S_t}{S_0}\right)-\frac{1}{\sigma_V}\log\left(\frac{V_t}{V_0}\right)}{\frac{\mu-\frac{1}{2}\sigma^2}{\sigma}-\frac{\mu_V-\frac{1}{2}\sigma_V^2}{\sigma_V}}R=Rt.$

We define the risk-free asset (designated as $\mathbb{B}$) with price process, $\mathscr{b}_t, t\geq 0$, having cumulative return in $[0,t]$

$$\log\left(\frac{\mathscr{b}_t}{\mathscr{b}_0}\right)=\frac{\frac{1}{\sigma}\log\left(\frac{S_t}{S_0}\right)-\frac{1}{\sigma_V}\log\left(\frac{V_t}{V_0}\right)}{\frac{\mu-\frac{1}{2}\sigma^2}{\sigma}-\frac{\mu_V-\frac{1}{2}\sigma_V^2}{\sigma_V}}R=Rt.$$

Because

$$\log\left(\frac{\mathscr{b}_t}{\mathscr{b}_0}\right)=\frac{\frac{\sigma_V}{\sigma_V-\sigma}\log\left(\frac{S_t}{S_0}\right)-\frac{\sigma}{\sigma_V-\sigma}\log\left(\frac{V_t}{V_0}\right)}{1+\frac{1}{2}\sigma\sigma_V(\frac{\sigma_V-\sigma}{\mu\sigma_V-\mu_V\sigma})}=\frac{\frac{\sigma}{\sigma-\sigma_V}\log\left(\frac{V_t}{V_0}\right)-\frac{\sigma_V}{\sigma-\sigma_V}\log\left(\frac{S_t}{S_0}\right)}{1+\frac{1}{2}\sigma\sigma_V(\frac{1}{R})}$$

and $\vartheta_V:=\frac{\sigma_V}{(\sigma-\sigma_V)\left[1+\frac{1}{2}\sigma\sigma_V(\frac{1}{R})\right]}$, $\vartheta:=\frac{\sigma}{(\sigma-\sigma_V)\left[1+\frac{1}{2}\sigma\sigma_V(\frac{1}{R})\right]}$

we have that $\mathscr{b}_t=\frac{{V_t}^{\vartheta}}{{S_t}^{\vartheta_V}}=\mathscr{b}_0e^{Rt}$.

Thus, the market with three assets that are publicly available for trade, $(S_t,V_t,\mathscr{b}_t)$ is free of arbitrage opportunities and complete.

Define then $\mathbb{Q}\sim\mathbb{P}$ on the stochastic basis $(\Omega,\mathcal{F},\{\mathcal{F}_t\}_{t\geq 0},\mathbb{Q})$, generated by the Brownian motion $B_t^{\mathbb{Q}}t\geq 0$, where on $\mathbb{P}$, $B_t^{\mathbb{Q}}=B_t+\vartheta t,\vartheta=\frac{\mu-R}{\sigma}$, where

$$\frac{\mu - R}{\sigma} = \frac{\mu - \frac{1}{\sigma - \sigma_V}(\mu_V\sigma - \mu\sigma_V)}{\sigma} = \frac{\mu\sigma - \mu_V\sigma}{\sigma(\sigma - \sigma_V)} = \frac{\mu - \mu_V}{\sigma - \sigma_V}$$

and

$$\frac{\mu_V - R}{\sigma_V} = \frac{\mu_V - \frac{1}{\sigma - \sigma_V}(\mu_V\sigma - \mu\sigma_V)}{\sigma_V} = \frac{\mu_V\sigma - \mu_V\sigma_V - (\mu_V\sigma - \mu\sigma_V)}{\sigma_V(\sigma - \sigma_V)} = \frac{\mu\sigma_V - \mu_V\sigma_V}{\sigma_V(\sigma - \sigma_V)} = \frac{\mu - \mu_V}{\sigma - \sigma_V}.$$

This completes the proof of Proposition 2.

***Proof of Proposition 3***

For simplicity of the exposition we consider the case $\mathrm{N} = 3$ only. Consider now a market with three risky assets (designated as $\mathbb{S}^{(j)}, j = 1,2,3$ ) with price dynamics the following three correlated GBMs:

$$dS_t^{(j)} = \mu^{(j)} S_t^{(j)} dt + S_t^{(j)}\left(\sigma^{(j)} dB_t + v^{(j)} dW_t\right), t \geq 0,\ S_0^{(j)} > 0, \mu^{(j)} > 0, \sigma^{(j)} > 0, v^{(j)} > 0$$

The no-arbitrage and market completeness conditions require that

$$\begin{bmatrix} \sigma^{(i)} & v^{(i)} \\ \sigma^{(j)} & v^{(j)} \end{bmatrix} \begin{bmatrix} \theta^{(1)} \\ \theta^{(2)} \end{bmatrix} = \begin{bmatrix} \mu^{(i)} - R \\ \mu^{(j)} - R \end{bmatrix}$$

holds for all $i, j = 1,2,3$. That is, $\sigma^{(j)}\theta^{(1)} + v^{(j)}\theta^{(2)} = \mu^{(j)} - R$ for all $j = 1,2,3$. Then,

$R = \mu^{(1)} - \sigma^{(1)}\theta^{(1)} - v^{(1)}\theta^{(2)}$ and $\left(\theta^{(1)}, \theta^{(2)}\right)^T$ satisfies the equation

$$\begin{bmatrix} \sigma^{(2)} & v^{(2)} \\ \sigma^{(3)} & v^{(3)} \end{bmatrix} \begin{bmatrix} \theta^{(1)} \\ \theta^{(2)} \end{bmatrix} = \begin{bmatrix} \mu^{(2)} - R \\ \mu^{(3)} - R \end{bmatrix}$$

This leads to

$$R=\frac{\left\{\begin{matrix}\mu^{(1)}\left(\sigma^{(2)}v^{(3)}-v^{(2)}\sigma^{(3)}\right)-\mu^{(2)}\left(\sigma^{(1)}v^{(3)}-v^{(1)}\sigma^{(3)}\right)+\\ +\mu^{(3)}\left(\sigma^{(1)}v^{(2)}-v^{(1)}\sigma^{(2)}\right)\end{matrix}\right\}}{\sigma^{(1)}\left(v^{(2)}-v^{(3)}\right)-\sigma^{(2)}\left(v^{(1)}-v^{(3)}\right)+\sigma^{(3)}\left(v^{(1)}-v^{(2)}\right)}$$

Next, by the Itô formula

$$dg\left(S_t^{(1)},S_t^{(2)},S_t^{(3)}\right)=\begin{bmatrix}\sum_{j=1}^{3}\frac{\partial g\left(S_t^{(1)},S_t^{(2)},S_t^{(3)}\right)}{\partial x^{(j)}}\,\mu^{(j)}S_t^{(j)}+\\ +\frac{1}{2}\sum_{j=1}^{3}\frac{\partial g^2\left(S_t^{(1)},S_t^{(2)},S_t^{(3)}\right)}{\partial x^{(j)^2}}\left(S_t^{(j)}\right)^2\left(\sigma^{(j)^2}+v^{(j)^2}\right)+\\ +\sum_{i=1}^{3}\sum_{j=i+1}^{3}\frac{\partial g^2\left(S_t^{(1)},S_t^{(2)},S_t^{(3)}\right)}{\partial x^{(i)}\partial x^{(j)}}\left(\sigma^{(i)}\sigma^{(j)}+v^{(i)}v^{(j)}\right)=\end{bmatrix}dt$$

$$+\sum_{j=1}^{3}\frac{\partial g\left(S_t^{(1)},S_t^{(2)},S_t^{(3)}\right)}{\partial x^{(j)}}\,S_t^{(j)}\sigma^{(j)}dB_t+\sum_{j=1}^{3}\frac{\partial g\left(S_t^{(1)},S_t^{(2)},S_t^{(3)}\right)}{\partial x^{(j)}}\,S_t^{(j)}v^{(j)}dW_t.$$

Consider the self-financing replicating portfolio

$$\mathrm{G_t}=g\left(S_t^{(1)},S_t^{(2)},S_t^{(3)}\right)=\sum_{j=1}^{3}a_t^{(j)}S_t^{(j)}.$$

Then,

$$\mathrm{dG_t}=dg\left(S_t^{(1)},S_t^{(2)},S_t^{(3)}\right)=\sum_{j=1}^{3}a_t^{(j)}dS_t^{(j)}=$$

$$=\sum_{j=1}^{3}a_t^{(j)}\mu^{(j)}S_t^{(j)}dt+\sum_{j=1}^{3}a_t^{(j)}\sigma^{(j)}S_t^{(j)}dB_t+\sum_{j=1}^{3}a_t^{(j)}v^{(j)}S_t^{(j)}dW_t\ .$$

Equating the terms for $\mathrm{dG_t}$ leads to:

$$a_t^{(1)}S_t^{(1)}=\frac{\begin{vmatrix}\sum_{j=1}^{3}\frac{\partial g\left(S_t^{(1)},S_t^{(2)},S_t^{(3)}\right)}{\partial x^{(j)}}S_t^{(j)}\sigma^{(j)}-\sigma^{(3)}g\left(S_t^{(1)},S_t^{(2)},S_t^{(3)}\right) & \left(\sigma^{(2)}-\sigma^{(3)}\right)\\ \sum_{j=1}^{3}\frac{\partial g\left(S_t^{(1)},S_t^{(2)},S_t^{(3)}\right)}{\partial x^{(j)}}S_t^{(j)}v^{(j)}-v^{(3)}g\left(S_t^{(1)},S_t^{(2)},S_t^{(3)}\right) & \left(v^{(2)}-v^{(3)}\right)\end{vmatrix}}{\begin{vmatrix}\left(\sigma^{(1)}-\sigma^{(3)}\right) & \left(\sigma^{(2)}-\sigma^{(3)}\right)\\ \left(v^{(1)}-v^{(3)}\right) & \left(v^{(2)}-v^{(3)}\right)\end{vmatrix}};$$

$$a_t^{(2)}S_t^{(2)} = \frac{\begin{vmatrix} \left(\sigma^{(1)}-\sigma^{(3)}\right) & \sum_{j=1}^{3}\frac{\partial g\left(S_t^{(1)},S_t^{(2)},S_t^{(3)}\right)}{\partial x^{(j)}}S_t^{(j)}\sigma^{(j)}-\sigma^{(3)}g\left(S_t^{(1)},S_t^{(2)},S_t^{(3)}\right) \\ \left(v^{(1)}-v^{(3)}\right) & \sum_{j=1}^{3}\frac{\partial g\left(S_t^{(1)},S_t^{(2)},S_t^{(3)}\right)}{\partial x^{(j)}}S_t^{(j)}v^{(j)}-v^{(3)}g\left(S_t^{(1)},S_t^{(2)},S_t^{(3)}\right) \end{vmatrix}}{\begin{vmatrix} \left(\sigma^{(1)}-\sigma^{(3)}\right) & \left(\sigma^{(2)}-\sigma^{(3)}\right) \\ \left(v^{(1)}-v^{(3)}\right) & \left(v^{(2)}-v^{(3)}\right) \end{vmatrix}};$$

and

$$a_t^{(3)}S_t^{(3)} = g\left(S_t^{(1)},S_t^{(2)},S_t^{(3)}\right) - \sum_{j=1}^{2} a_t^{(j)}S_t^{(j)} =$$

$$= g\left(S_t^{(1)},S_t^{(2)},S_t^{(3)}\right) -$$

$$- \frac{\begin{vmatrix} \sum_{j=1}^{3}\frac{\partial g\left(S_t^{(1)},S_t^{(2)},S_t^{(3)}\right)}{\partial x^{(j)}}S_t^{(j)}\sigma^{(j)}-\sigma^{(3)}g\left(S_t^{(1)},S_t^{(2)},S_t^{(3)}\right) & \left(\sigma^{(2)}-\sigma^{(3)}\right) \\ \sum_{j=1}^{3}\frac{\partial g\left(S_t^{(1)},S_t^{(2)},S_t^{(3)}\right)}{\partial x^{(j)}}S_t^{(j)}v^{(j)}-v^{(3)}g\left(S_t^{(1)},S_t^{(2)},S_t^{(3)}\right) & \left(v^{(2)}-v^{(3)}\right) \end{vmatrix}}{\begin{vmatrix} \left(\sigma^{(1)}-\sigma^{(3)}\right) & \left(\sigma^{(2)}-\sigma^{(3)}\right) \\ \left(v^{(1)}-v^{(3)}\right) & \left(v^{(2)}-v^{(3)}\right) \end{vmatrix}} -$$

$$- \frac{\begin{vmatrix} \left(\sigma^{(1)}-\sigma^{(3)}\right) & \sum_{j=1}^{3}\frac{\partial g\left(S_t^{(1)},S_t^{(2)},S_t^{(3)}\right)}{\partial x^{(j)}}S_t^{(j)}\sigma^{(j)}-\sigma^{(3)}g\left(S_t^{(1)},S_t^{(2)},S_t^{(3)}\right) \\ \left(v^{(1)}-v^{(3)}\right) & \sum_{j=1}^{3}\frac{\partial g\left(S_t^{(1)},S_t^{(2)},S_t^{(3)}\right)}{\partial x^{(j)}}S_t^{(j)}v^{(j)}-v^{(3)}g\left(S_t^{(1)},S_t^{(2)},S_t^{(3)}\right) \end{vmatrix}}{\begin{vmatrix} \left(\sigma^{(1)}-\sigma^{(3)}\right) & \left(\sigma^{(2)}-\sigma^{(3)}\right) \\ \left(v^{(1)}-v^{(3)}\right) & \left(v^{(2)}-v^{(3)}\right) \end{vmatrix}}.$$

Equating the terms for $dt$ leads to the BSM-equation in three dimensions:

$$\sum_{j=1}^{3}\frac{\partial g\left(S_t^{(1)},S_t^{(2)},S_t^{(3)}\right)}{\partial x^{(j)}}\,RS_t^{(j)} - Rg\left(S_t^{(1)},S_t^{(2)},S_t^{(3)}\right) +$$
$$+\frac{1}{2}\sum_{j=1}^{3}\frac{\partial g^2\left(S_t^{(1)},S_t^{(2)},S_t^{(3)}\right)}{\partial {x^{(j)}}^2}\left(S_t^{(j)}\right)^2\left({\sigma^{(j)}}^2+{v^{(j)}}^2\right) +$$
$$+\sum_{i=1}^{3}\sum_{j=i+1}^{3}\frac{\partial g^2\left(S_t^{(1)},S_t^{(2)},S_t^{(3)}\right)}{\partial x^{(i)}\partial x^{(j)}}\left(\sigma^{(i)}\sigma^{(j)}+v^{(i)}v^{(j)}\right) = 0$$

This proves Proposition 3 in the case of $N = 3$. The general case is considered in the same manner.

***Proof of Proposition 4***

By the Itô formula

$$dg\left(\mathcal{S}_t^{(B)}\right) = \left\{\begin{array}{c} \frac{\partial g\left(\mathcal{S}_t^{(B)}\right)}{\partial x^{(1)}} S_t^{(1,B)}\mu^{(1)} + \frac{\partial g\left(\mathcal{S}_t^{(B)}\right)}{\partial x^{(2)}} S_t^{(2,B)}\mu^{(2)} + \frac{\partial g\left(\mathcal{S}_t^{(B)}\right)}{\partial x^{(3)}} S_t^{(3,B)}\mu^{(3)} + \\ +\frac{1}{2}\frac{\partial^2 g\left(\mathcal{S}_t^{(B)}\right)}{\partial {x^{(1)}}^2}\left(\sigma^{(1)} S_t^{(1,B)}\right)^2 + \frac{1}{2}\frac{\partial^2 g\left(\mathcal{S}_t^{(B)}\right)}{\partial {x^{(2)}}^2}\left(\sigma^{(2)} S_t^{(2,,B)}\right)^2 + \\ + \frac{\partial g\left(\mathcal{S}_t^{(B)}\right)}{\partial x^{(1)}\partial x^{(2)}} S_t^{(1,B)} S_t^{(2,B)}\sigma^{(1)}\sigma^{(2)} \end{array}\right\} dt \; +$$

$$+\left\{\frac{\partial g\left(\mathcal{S}_t^{(B)}\right)}{\partial x^{(1)}} S_t^{(1,B)}\sigma^{(1)} + \frac{\partial g\left(\mathcal{S}_t^{(B)}\right)}{\partial x^{(1)}} S_t^{(2,,B)}\sigma^{(2)}\right\} dB_t \; +$$

$$+\left\{\begin{array}{c} \frac{\partial g\left(\mathcal{S}_t^{(B)}\right)}{\partial x^{(1)}} S_t^{(1,B)}\gamma + \frac{\partial g\left(\mathcal{S}_t^{(B)}\right)}{\partial x^{(3)}} S_t^{(3,B)}\gamma + \\ +\frac{1}{2}\frac{\partial^2 g\left(\mathcal{S}_t^{(B)}\right)}{\partial {x^{(1)}}^2}\left(\gamma S_t^{(1,B)}\right)^2 + \frac{1}{2}\frac{\partial^2 g\left(\mathcal{S}_t^{(B)}\right)}{\partial {x^{(3)}}^2}\left(\gamma S_t^{(3,,B)}\right)^2 + \\ + \frac{\partial g\left(\mathcal{S}_t^{(B)}\right)}{\partial x^{(1)}\partial x^{(3)}} S_t^{(1,B)} S_t^{(3,,B)}\gamma^2 \end{array}\right\} dN_t$$

Consider a self-financing strategy, $g\left(\mathcal{S}_t^{(B)}\right) = a_t^{(1)} S_t^{(1,B)} + a_t^{(2)} S_t^{(2,B)} + a_t^{(3)} S_t^{(3,,B)}$, and thus $dg\left(\mathcal{S}_t^{(B)}\right) = a_t^{(1)} dS_t^{(1,B)} + a_t^{(2)} dS_t^{(2,B)} + a_t^{(3)} dS_t^{(3,,B)}$. Comparing the terms for $dg\left(\mathcal{S}_t^{(B)}\right)$ leads to

$$a_t^{(1)} S_t^{(1,B)} = \frac{1}{\sigma^{(1)}}\left(A_t - \left(g\left(\mathcal{S}_t^{(B)}\right) - \frac{B_t}{\gamma}\right)\sigma^{(2)}\right), a_t^{(2)} S_t^{(2,B)} = \; g\left(\mathcal{S}_t^{(B)}\right) - \frac{B_t}{\gamma},$$

$$a_t^{(3)} S_t^{(3,,B)} = \frac{B_t}{\gamma} - \frac{1}{\sigma^{(1)}}\left(A_t - \left(g\left(\mathcal{S}_t^{(B)}\right) - \frac{B_t}{\gamma}\right)\sigma^{(2)}\right),$$

where $A_t = \frac{\partial g\left(\mathcal{S}_t^{(B)}\right)}{\partial x^{(1)}} S_t^{(1,B)} \sigma^{(1)} + \frac{\partial g\left(\mathcal{S}_t^{(B)}\right)}{\partial x^{(1)}} S_t^{(2,,B)} \sigma^{(2)}$ and

$$B_t = \frac{\partial g\left(\mathcal{S}_t^{(B)}\right)}{\partial x^{(1)}} S_t^{(1,B)} \gamma + \frac{\partial g\left(\mathcal{S}_t^{(B)}\right)}{\partial x^{(3)}} S_t^{(3,B)} \gamma + \frac{1}{2}\frac{\partial^2 g\left(\mathcal{S}_t^{(B)}\right)}{\partial {x^{(1)}}^2}\left(\gamma S_t^{(1,B)}\right)^2 + \frac{1}{2}\frac{\partial^2 g\left(\mathcal{S}_t^{(B)}\right)}{\partial {x^{(3)}}^2}\left(\gamma S_t^{(3,,B)}\right)^2$$

$$+ \frac{\partial g\left(\mathcal{S}_t^{(B)}\right)}{\partial x^{(1)} \partial x^{(3)}} S_t^{(1,B)} S_t^{(3,,B)} \gamma^2.$$

Then equating the terms with $dt$ results in

$$\frac{\partial g\left(\mathcal{S}_t^{(B)}\right)}{\partial x^{(1)}} S_t^{(1,B)} \mu^{(1)} + \frac{\partial g\left(\mathcal{S}_t^{(B)}\right)}{\partial x^{(2)}} S_t^{2(,B)} \mu^{(2)} + \frac{\partial g\left(\mathcal{S}_t^{(B)}\right)}{\partial x^{(3)}} S_t^{(,B)} \mu^{(3)} +$$

$$+ \frac{1}{2}\frac{\partial^2 g\left(\mathcal{S}_t^{(B)}\right)}{\partial {x^{(1)}}^2}\left(\sigma^{(1)} S_t^{(1,B)}\right)^2 + \frac{1}{2}\frac{\partial^2 g\left(\mathcal{S}_t^{(B)}\right)}{\partial {x^{(2)}}^2}\left(\sigma^{(2)} S_t^{(2,,B)}\right)^2 +$$

$$+ \frac{\partial g\left(\mathcal{S}_t^{(B)}\right)}{\partial x^{(1)} \partial x^{(2)}} S_t^{(1,B)} S_t^{(2,B)} \sigma^{(1)} \sigma^{(2)} =$$

$$= a_t^{(1)} S_t^{(1,B)} \mu^{(1)} + a_t^{(2)} S_t^{(2,B)} \mu^{(2)} + a_t^{(3)} S_t^{(3,B)} \mu^{(3)}.$$

Thus,

$$R\frac{\partial g\left(\mathcal{S}_t^{(B)}\right)}{\partial x^{(1)}} S_t^{(1,B)} + R\frac{\partial g\left(\mathcal{S}_t^{(B)}\right)}{\partial x^{(2)}} S_t^{2(,B)} + R\frac{\partial g\left(\mathcal{S}_t^{(B)}\right)}{\partial x^{(3)}} S_t^{(,B)} - Rg\left(\mathcal{S}_t^{(B)}\right) +$$

$$+ \frac{1}{2}\frac{\partial^2 g\left(\mathcal{S}_t^{(B)}\right)}{\partial {x^{(1)}}^2}\left(\sigma^{(1)} S_t^{(1,B)}\right)^2 + \frac{1}{2}\frac{\partial^2 g\left(\mathcal{S}_t^{(B)}\right)}{\partial {x^{(2)}}^2}\left(\sigma^{(2)} S_t^{(2,,B)}\right)^2 + \frac{\partial g\left(\mathcal{S}_t^{(B)}\right)}{\partial x^{(1)} \partial x^{(2)}} S_t^{(1,B)} S_t^{(2,B)} \sigma^{(1)} \sigma^{(2)} +$$

$$+(R-\mu^{(3)})\gamma\left\{\begin{array}{c}+\frac{1}{2}\frac{\partial^2 g\left(\mathcal{S}_t^{(B)}\right)}{\partial {x^{(1)}}^2}\left(S_t^{(1,B)}\right)^2+\frac{1}{2}\frac{\partial^2 g\left(\mathcal{S}_t^{(B)}\right)}{\partial {x^{(3)}}^2}\left(S_t^{(3,,B)}\right)^2+\\ +\frac{\partial g\left(\mathcal{S}_t^{(B)}\right)}{\partial x^{(1)}\partial x^{(3)}}S_t^{(1,B)}S_t^{(2,,B)}\end{array}\right\}=0,$$

where $R=\frac{\mu^{(2)}\sigma^{(1)}-\sigma^{(2)}\mu^{(1)}+\mu^{(3)}\sigma^{(2)}}{\sigma^{(1)}}$. This completes the proof of Proposition 4.

***Proof of Proposition 6***

Applying fractional Stratonovich integration (pathwise integration)

$$dG_t=dg(S_t,V_t)=\frac{\partial g(S_t,V_t)}{\partial x}dS_t+\frac{\partial g(S_t,V_t)}{\partial y}dV_t=$$

$$=\frac{\partial g(S_t,V_t)}{\partial x}S_t\left(\left(2B_t^{(H)}\mu+\sigma\right)*dB_t^{(H)}\right)+\frac{\partial g(S_t,V_t)}{\partial y}V_t\left(\left(2B_t^{(H)}\mu_V+\sigma_V\right)*dB_t^{(H)}\right)+$$

$$=\left\{\left\{2\mu\frac{\partial g(S_t,V_t)}{\partial x}S_t+2\mu_V\frac{\partial g(S_t,V_t)}{\partial y}V_t\right\}B_t^{(H)}+\sigma\frac{\partial g(S_t,V_t)}{\partial x}S_t+\sigma_V\frac{\partial g(S_t,V_t)}{\partial y}V_t\right\}*dB_t^{(H)}.$$

Consider a self-financing replicating strategy: $g(S_t,V_t)=a_tS_t+b_tV_t$.Thus,

$$dg(S_t,V_t)=a_tdS_t+b_tdV_t=\left(a_tS_t\left(2\mu B_t^{(H)}+\sigma\right)+b_tV_t\left(2\mu_VB_t^{(H)}+\sigma_V\right)\right)*dB_t^{(H)}=$$

$$=\left([2\mu a_tS_t+2\mu_Vb_tV_t]B_t^{(H)}+a_tS_t\sigma+b_tV_t\sigma_V\right)*dB_t^{(H)}.$$

Equating the terms for $dg(S_t,V_t)$ leads to:

$$\left\{2\mu\frac{\partial g(S_t,V_t)}{\partial x}S_t+2\mu_V\frac{\partial g(S_t,V_t)}{\partial y}V_t\right\}B_t^{(H)}+\sigma\frac{\partial g(S_t,V_t)}{\partial x}S_t+\sigma_V\frac{\partial g(S_t,V_t)}{\partial y}V_t=[2\mu a_tS_t+$$
$$2\mu_Vb_tV_t]B_t^{(H)}+a_tS_t\sigma+b_tV_t\sigma_V.$$

Therefore,

$$a_t S_t = \frac{\mu\sigma_V \frac{\partial g(S_t,V_t)}{\partial x} S_t + \mu_V\sigma_V \frac{\partial g(S_t,V_t)}{\partial y} V_t - \mu_V\sigma \frac{\partial g(S_t,V_t)}{\partial x} S_t + \mu_V\sigma_V \frac{\partial g(S_t,V_t)}{\partial y} V_t}{\mu\sigma_V - \mu_V\sigma},$$

and

$$b_t V_t = \frac{\mu\sigma \frac{\partial g(S_t,V_t)}{\partial x} S_t + \mu\sigma_V \frac{\partial g(S_t,V_t)}{\partial y} V_t - \mu\sigma \frac{\partial g(S_t,V_t)}{\partial x} S_t - \mu_V\sigma \frac{\partial g(S_t,V_t)}{\partial y} V_t}{\mu\sigma_V - \mu_V\sigma}.$$

Applying the equality $g(S_t, V_t) = a_t S_t + b_t V_t$ completes the proof of Proposition 6.*Proof of*

***Proof of Proposition 7***

By the pathwise formula

$$g(\mathcal{S}_t) = \sum_{k=1}^{N-1} \left\{ \begin{array}{c} \left( 2\sum_{j=1}^{N} \frac{\partial g(\mathcal{S}_t)}{\partial x^{(j)}} \mu^{(j)} S_t^{(j)} \sum_{m=1}^{N-1} B_t^{(H,m)} \right) + \\ + \left( \sum_{j=1}^{N} \frac{\partial g(\mathcal{S}_t)}{\partial x^{(j)}} S_t^{(j)} \sigma^{(j,k)} \right) \end{array} \right\} \circledast dB_t^{(H,k)}$$

Consider the replication self-financing strategy: $g(\mathcal{S}_t) = \sum_{j=1}^{N} a_t^{(j)} S_t^{(j)}$ with

$$dg(\mathcal{S}_t) = \sum_{j=1}^{N} a_t^{(j)} dS_t^{(j)} = \sum_{j=1}^{N} a_t^{(j)} S_t^{(j)} \left( \sum_{k=1}^{N-1} \left( 2\mu^{(j)} \sum_{m=1}^{N-1} B_t^{(H,m)} \right) \circledast dB_t^{(H,k)} \right)$$

$$+ \sum_{k=1}^{N-1} \left( \sum_{j=1}^{N} a_t^{(j)} S_t^{(j)} \sigma^{(j,k)} \right) \circledast dB_t^{(H,k)}$$

Comparing the terms for $dg(\mathcal{S}_t)$ leads to:

$$\sum_{j=1}^{N} \frac{\partial g(\mathcal{S}_t)}{\partial x^{(j)}} \mu^{(j)} S_t^{(j)} = \sum_{j=1}^{N} a_t^{(j)} S_t^{(j)} \mu^{(j)},$$

$$\sum_{j=1}^{N}\frac{\partial g(\mathcal{S}_t)}{\partial x^{(j)}}S_t^{(j)}\sigma^{(j,k)} = \sum_{j=1}^{N} a_t^{(j)}S_t^{(j)}\sigma^{(j,k)}, k = 1, \ldots, N-1.$$

Consider the linear system

$$\sum_{j=1}^{N}\mu^{(\mathrm{j})}\,a_t^{(j)}S_t^{(j)} = \sum_{j=1}^{N}\frac{\partial g(\mathcal{S}_t)}{\partial x^{(j)}}\mu^{(\mathrm{j})}S_t^{(j)}\ ,\sum_{j=1}^{N}\sigma^{(j,k)}\,a_t^{(j)}S_t^{(j)} = \sum_{j=1}^{N}\frac{\partial g(\mathcal{S}_t)}{\partial x^{(j)}}S_t^{(j)}\sigma^{(j,k)}$$

with unknows $a_t^{(j)}, j = 1, .., N$. Because by assumption $det\Xi \neq 0$, the system has a unique solution given by $a_t^{(j)} = \frac{\partial g(\mathcal{S}_t)}{\partial x^{(j)}}, j = 1, \ldots, N$. Thus, the linear PDE for the perpetual derivative is given by $\sum_{j=1}^{N} \frac{\partial g(\mathcal{S}_t)}{\partial x^{(j)}}S_t^{(j)} - g(\mathcal{S}_t) = 0$, completing the proof of Proposition 7.